\documentclass[pra,aps,showpacs,preprint]{revtex4}
\usepackage{graphicx}
\usepackage{amsmath}

\begin{document}
\title{\bf Quantum Teleportation through Noisy Channels with Multi-Qubit GHZ States}
\author{Pakhshan Espoukeh}
\author{Pouria Pedram}
\email[Electronic address: ]{p.pedram@srbiau.ac.ir}
\affiliation{Department of Physics, Tehran Science and Research
Branch, Islamic Azad University, Tehran, Iran}
\date{\today}

\begin{abstract}
We investigate two-party quantum teleportation through noisy
channels for multi-qubit Greenberger-Horne-Zeilinger (GHZ) states and find
which state loses less quantum information in the process. The
dynamics of states is described by the master equation with the
noisy channels that lead to the quantum channels to be mixed states.
We analytically solve the Lindblad equation for $n$-qubit GHZ states
$n\in\{4,5,6\}$ where Lindblad operators correspond to the Pauli
matrices and describe the decoherence of states. Using the average
fidelity we show that 3GHZ state is more robust than $n$GHZ state
under most noisy channels. However, $n$GHZ state preserves same
quantum information with respect to EPR and 3GHZ states where the
noise is in $x$ direction in which the fidelity remains unchanged.
We explicitly show that Jung \emph{et al.} conjecture [Phys. Rev. A {\bf 78}, 012312 (2008)], namely,
``average fidelity with same-axis noisy channels are in general
larger than average fidelity with different-axis noisy channels'' is
not valid for 3GHZ and 4GHZ states.
\end{abstract}

 \maketitle

\section{Introduction}\label{sec1}
Quantum teleportation is a process based on classical communication
that transmits the quantum information from a location to another
with the help of shared quantum entanglement between the sender and
receiver. This process is a technique for transporting the state of
an atom or photon to the remote recipient even in the absence of
quantum communication channels connecting the sender of the quantum
state (called  Alice) to the recipient (called Bob) \cite{nel}. The
original protocol of this process was firstly introduced by Bennett
\emph{et al.} using the Einstein-Podolsky-Rosen (EPR) state as the
quantum channel \cite{Bennett93}. Quantum teleportation using two
qubit systems is discussed also in \cite{chakra10,liang13}.

Because of the strong connection between quantum entanglement and
quantum teleportation, the usage of multiparticle entangled quantum
states other than two-particle entangled states for quantum
teleportation has been the subject of various investigations
\cite{wang07,paul11}. In particular, quantum teleportation with
three-qubit GHZ state and W state is studied in
Refs.~\cite{karl98,wang10,gorb03,joo03,agra06,Hu1}. The
possibility of teleportation of an unknown qubit using four-particle
GHZ state is discussed in Ref.~\cite{pati00}. It is also shown that
the state in the form $|{\psi}\rangle = \frac{1}{\sqrt{2}} \left(
|00q_1\rangle + |11q_2\rangle \right)$ allows perfect two-party
teleportation in which $|q_1\rangle$ and $|q_2\rangle$ are arbitrary
normalized single qubit states \cite{jung07}.

In quantum information theory and quantum computation, fidelity is a
measure to quantify the closeness of two quantum states \cite{nel} and
is closely related to quantum entanglement \cite{4}, quantum phase
transitions \cite{6,7,8}, and quantum chaos \cite{5}. Fidelity can
be also used to quantify how much quantum information is lost due to
noisy channel between initial and final states. This reduction of fidelity is usually due to
the interaction of quantum states with environment which results in
imperfect teleportation. Thus, the coherence of the entangled state
may be lost and it becomes a mixed state. Some efforts have been
performed in this direction to realize effective factors which cause
this phenomenon \cite{Bennett93,horodecki99,banas01,jung08-1}. For
instance, Bennett \emph{et al.} showed that the fidelity of
teleportation and the range of accurately teleported states reduce in
the less entangled quantum channels \cite{Bennett93}.

The existence of noise is an unavoidable property in quantum
teleportation process which results is decoherence of states and the
reduction of fidelity \cite{0h02,han08,bhak08}. In particular, Oh
\emph{et al.} using a pair of EPR states showed that the average
fidelity and the range of teleported states depend on the type of
the noise that acts on the quantum channel and confirmed Bennett
\emph{et al.} results \cite{0h02}. They solved analytically and
numerically the master equation with Lindblad structure and found
the fidelity as a function of decoherence time and angles of an
unknown teleported state.

Note that analytically solving the Lindblad equation in the presence of the noise is
not a trivial task in general. Indeed, for multiparticle systems one
needs to solve many coupled differential equations that involve
tedious computation. For example, for three-particle GHZ state
(3GHZ) the master equation reduces to 8 diagonal coupled
differential equations and 28 off-diagonal coupled differential
equations \cite{jung08-2}. The situation is even worse for
four-particle GHZ state (4GHZ) that involves 16 diagonal coupled
differential equations and 120 off-diagonal coupled differential
equations.

In this paper, we analytically solve the master equation for
$n$-particle GHZ state ($n\in\{4,5,6\}$) through various noisy
channels. The number of coupled differential equations for each case
is considerably reduced by using a proper ansatz for the density
matrix.  The ansatz is determined from the temporal evolution of the
initial state of the system. We obtain the fidelity of teleportation
and the average fidelity of teleportation that depend on the type of
the noisy channel and compare the results with three-particle GHZ
state. The goal of this paper is to find out which state is better
(loses less quantum information) in the teleportation process with
noisy channels. Therefore, although various noisy channels were
studied in Ref.~\cite{0h02}, we discuss noisy channels which cause
the quantum channels to be mixed to compare $n$GHZ states in the
process of teleportation.

The organization of this paper is as follows: Section \ref{sec2} is
devoted to general framework used to evaluate the two-party quantum
teleportation circuit. In Sec.~\ref{sec3}, we analytically solve the
Lindblad equation where the quantum channel is a four-particle GHZ
state, i.e., $| \mbox{4GHZ}\rangle$. We transmit 4GHZ state through
isotropic and Pauli noises and compute the fidelity of
teleportation. Moreover, we compare the robustness of 4GHZ state
with 3GHZ state in the noisy channels. Solving the master equation
for 5GHZ and 6GHZ states when Lindblad operators are in $x$ and $z$
directions is the subject of Secs.~\ref{sec4} and \ref{sec5},
respectively. We present our conclusions in Sec.~\ref{sec6}.

\begin{figure}[t]
%
%
\begin{picture}(200,120)
\put(1,59){\makebox(10,10){$|\psi_{\text{in}}\rangle$}}
\put(-2.5,25){\makebox(10,12){$|\mbox{EPR}\rangle$}}
\put(20,28){$\Biggl\{$}

\put(32.5,15){\dashbox{1}(12.5,35){}}

\put(30,65){\line(1,0){45}}\put(75,65){\line(1,0){15}}\put(105,65){\line(1,0){80}}
\put(30,42.5){\line(1,0){154}}
\put(30,20){\line(1,0){97.5}}\put(142.5,20){\line(1,0){15}}\put(172.5,20){\line(1,0){12.5}}

\put(65,65){\circle*{4}} \put(65,66){\line(0,-1){26.25}}

\put(65,42.5){\circle{5}}

\put(90,57.5){\framebox(15,15){H}}

\put(135,42.5){\circle*{4}}
\put(135,42.5){\line(0,-1){15}}
\put(127.5,12.5){\framebox(15,15){$X$}}

\put(165,65){\circle*{4}}
\put(165,65){\line(0,-1){37.5}}
\put(157.5,12.5){\framebox(15,15){$Z$}}

\put(190,60){\makebox(10,10){$M$}}
\put(190,37.5){\makebox(10,10){$M$}}
\put(200,15){\makebox(10,10){$|\psi_{\text{out}}\rangle$ }}
\end{picture}
\caption{\label{fig1}A circuit for quantum teleportation through
noisy channels with EPR state. The two top lines belong to Alice and
the bottom line to Bob. $M$ denotes measurement and the dotted box
represents noisy channel. The Lindblad operator is turned on inside
the dotted box.}
\end{figure}
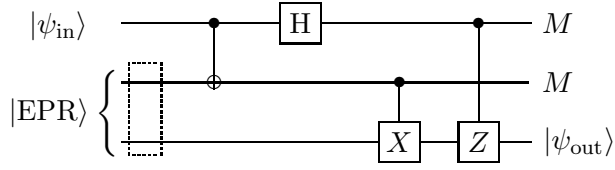

\section{GHZ state, fidelity, and Lindblad equation}\label{sec2}
For $n$-particle system, an $n$GHZ state is a quantum state defined
as follows
\begin{equation}
|n\mbox{GHZ}\rangle=\frac{1}{\sqrt{2}} \left( |0\rangle^{\otimes n}
+ |1\rangle^{\otimes n} \right),
\end{equation}
where $n>2$. Note that, teleportation with $|\mbox{EPR}\rangle$ through
noisy channels is depicted in Fig.~\ref{fig1} and it is discussed in
Ref.~\cite{0h02}. Also, teleportation of 3GHZ state through various
noisy channels has been previously studied in Ref.~\cite{jung08-2}.
Here, we are interested to investigate the teleportation process for
$n$GHZ state through noisy channels for $n\in\{4,5,6\}$. For this
purpose, we need to solve the master equation with Lindblad form
\cite{lind76}
\begin{equation}\label{Lindblad}
\frac{\partial \rho}{\partial t} = -\frac{i}{\hbar} [H_S, \rho] +
\sum_{i, \alpha} \left(L_{i,\alpha} \rho L_{i,\alpha}^{\dagger} -
\frac{1}{2} \left\{ L_{i,\alpha}^{\dagger} L_{i,\alpha}, \rho
\right\} \right),
\end{equation}
in which $L_{i,\alpha} = \sqrt{\kappa_{i,\alpha}}
\sigma^{(i)}_{\alpha}$ denote Lindblad operators that describe
decoherence and act on the $i$th qubit. Also,
$\sigma^{(i)}_{\alpha}$ are the Pauli spin matrices of the $i$th
qubit with $\alpha =\{ x,y,z\}$, $\kappa_{i,\alpha}$ is the
decoherence rate, and $H_S$ is the Hamiltonian of the system.

\begin{figure}
%
%
\begin{picture}(200,100)
\put(0,94){\makebox(10,10){$|\psi_{\text{in}}\rangle$}}
\put(-15,44){\makebox(10,10){$|4\mbox{GHZ}\rangle$}}
\put(10,44){\huge$\Biggl\{$}

\put(32.5,15){\dashbox{1}(12.5,70){}}

\put(30,100){\line(1,0){85}}\put(130,100){\line(1,0){50}}
\put(30,80){\line(1,0){150}}
\put(30,60){\line(1,0){150}}
\put(30,40){\line(1,0){150}}
\put(30,20){\line(1,0){107.5}}\put(152.5,20){\line(1,0){5}}\put(172.5,20){\line(1,0){7.5}}

\put(60,100){\circle*{4}}
\put(60,100){\line(0,-1){22.5}}
\put(60,80){\circle{5}}

\put(80,100){\circle*{4}}
\put(80,100){\line(0,-1){42.5}}
\put(80,60){\circle{5}}

\put(100,100){\circle*{4}}
\put(100,100){\line(0,-1){62.5}}
\put(100,40){\circle{5}}

\put(115,92.5){\framebox(15,15){H}}

\put(145,40){\circle*{4}}
\put(145,40){\line(0,-1){12.5}}
\put(137.5,12.5){\framebox(15,15){$X$}}

\put(165,100){\circle*{4}}
\put(165,100){\line(0,-1){72.5}}
\put(157.5,12.5){\framebox(15,15){$Z$}}

\put(190,95){\makebox(10,10){$M$}}
\put(190,75){\makebox(10,10){$M$}}
\put(190,55){\makebox(10,10){$M$}}
\put(190,35){\makebox(10,10){$M$}}
\put(200,15){\makebox(10,10){$|\psi_{\text{out}}\rangle$ }}
\end{picture}
\caption{\label{fig2}
   A circuit for quantum teleportation through noisy channels with 4GHZ state.
   The four top lines belong to Alice and the bottom line to Bob.
   $M$ denotes measurement and the dotted box represents noisy channel.
   The Lindblad operator is turned on inside the dotted box.}
\end{figure}
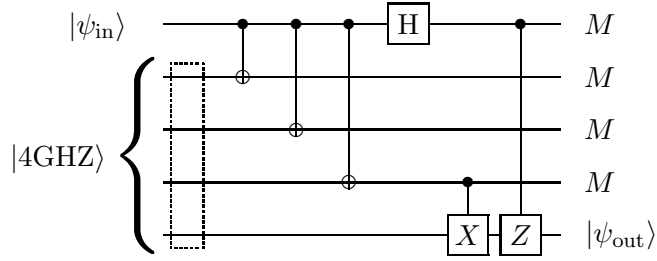

The unknown state to be teleportated can be written as a Bloch vector on a
Bloch sphere
\begin{equation}
\label{unknown} |\psi_{\text{in}}\rangle = \cos
\left(\frac{\theta}{2}\right) e^{i \phi / 2} |0\rangle + \sin
\left(\frac{\theta}{2}\right) e^{-i \phi / 2} |1\rangle,
\end{equation}
where $\theta$ and $\phi$ denote the polar and azimuthal angles,
respectively. Fig.~\ref{fig2} shows a quantum teleportation circuit
through noisy channels with 4GHZ state in which the input state
involves five qubits as the product state of
$|\psi_{\text{in}}\rangle $ and $|\mbox{4GHZ}\rangle$. The four top
lines (qubits) belong to Alice and bottom one belongs to Bob. The
difference of this circuit with the teleportation circuit for EPR
state (Fig.~\ref{fig1}) is the presence of two more controlled-NOT
$\left(\mbox{CNOT} \right)$ gates between $\psi_{\text{in}}$ and
$\mbox{4GHZ}$ states. After measurement of the top four qubits, Bob
gets the teleported state $|\psi_{\text{out}}\rangle$. It is
convenient to describe the teleportation in terms of the density
operator
\begin{equation}
\label{out1} \rho_{\text{out}} = \mbox{Tr}_{1,2,3,4} \left[
U_{\mbox{\tiny tel}} \rho_{in} \otimes \varepsilon
(\rho_{4\mbox{\tiny GHZ}}) U_{\mbox{\tiny tel}}^{\dagger} \right],
\end{equation}
where $\rho_{\text{in}} = |\psi_{\text{in}}\rangle \langle
\psi_{\text{in}}|$ is density matrix of the unknown initial state
and $\varepsilon(\rho_{4\mbox{\tiny GHZ}})$ is the density matrix
after transmission through noisy channel which is given by the
Lindblad equation. In fact, $\varepsilon$ is a quantum operation
that maps $\rho_{4\mbox{\tiny GHZ}}$ to $\varepsilon
(\rho_{4\mbox{\tiny GHZ}})$ because of noisy channel and
$\rho_{4\mbox{\tiny GHZ}}=|\mbox{4GHZ}\rangle\langle\mbox{4GHZ}|$.
Moreover, $U_{\mbox{\tiny tel}}$ is the unitary operator
corresponding to the quantum circuit and $\mbox{Tr}_{1,2,3,4}$ is
partial trace over first four qubits which belong to Alice.

Fidelity can be used as a tool to measure how much information is
lost or preserved through noisy quantum channels in quantum
teleportation process. It can be written as the overlap between the
input state $|\psi_{in}\rangle$ and the density operator for the
teleported state $|\rho_{\text{out}} \rangle$,
\begin{equation}
\label{fidelity} F = \langle \psi_{\text{in}} | \rho_{\text{out}} |
\psi_{\text{in}}\rangle,
\end{equation}
that depends on an input state and the type of noise. For the
perfect teleportation the fidelity is equal to unity. Also, $1-F$
indicates how much information is lost through the teleportation
process. For all possible unknown input states,
the average fidelity is given by
\begin{equation}
\overline{}\label{average} \overline{F} = \frac{1}{4 \pi}
\int_{0}^{2 \pi} \mathrm{d} \phi \int_{0}^{\pi} \mathrm{d} \theta
\sin \theta F(\theta, \phi).
\end{equation}
Similarly, we find the unitary operator, fidelity and average
fidelity for 5GHZ and 6GHZ states in the following sections.

\section{Four-qubit GHZ state with noisy channels}\label{sec3}
In this section, we analytically solve the Lindblad equation,
Eq.~(\ref{Lindblad}), for 4GHZ state through various noisy channels.
First, consider $(L_{2,x},L_{3,x},L_{4,x},L_{5,x})$ noise channel
with $\kappa_{2,x}=\kappa_{3,x}=\kappa_{4,x}=\kappa_{5,x}=\kappa$
that acts on 4GHZ state. Also, here and throughout the paper we
assume $H_S=0$.

For this case, the Lindblad equation involves 16 diagonal and 120
off-diagonal coupled linear differential equations which make this
equation difficult to be solved analytically. To overcome this problem,
we find the time evolution of the density matrix for infinitesimal
time interval $\delta t$ using the Lindblad equation as
\begin{equation}\label{del}
\rho(\delta t)= \rho(0) + \left[ \sum_{i, \alpha} \left(L_{i,\alpha}
\rho(0) L_{i,\alpha}^{\dagger} \right) - \frac{1}{2} \left\{
L_{i,\alpha}^{\dagger} L_{i,\alpha}, \rho(0) \right\} \right]\delta
t,
\end{equation}
where
\begin{equation}\label{ro0}
\rho(0)=|\mbox{4GHZ}\rangle\langle\mbox{4GHZ}|=\frac{1}{2}\left[|0\rangle^{\otimes
4} \langle 0 |^{\otimes 4} + |0\rangle^{\otimes 4}\langle
1|^{\otimes 4}+|1\rangle^{\otimes 4} \langle 0 |^{\otimes 4} +
|1\rangle ^{\otimes 4}\langle 1|^{\otimes 4}\right].
\end{equation}
Substituting $\rho(0)$ in Eq.~(\ref{del}) results in
\begin{eqnarray}
\varepsilon(\rho_{4\mbox{\tiny GHZ}})\Big|_{t=\delta t} =
\frac{1}{2}{\left(
\begin{smallmatrix}
1 -4 \kappa \delta t & 0 & 0 & 0 & 0 & 0 &0 & 0 & 0 & 0 & 0 & 0 & 0 & 1-4 \kappa \delta t   \\
0 & \kappa \delta t & 0 & 0& 0 & 0 & 0& 0 & 0 & 0 & 0 & 0 &  \kappa \delta t & 0   \\
0 & 0 & \kappa \delta t & 0& 0 & 0& 0 & 0 & 0 & 0 & 0 & \kappa \delta t & 0 & 0   \\
0 & 0 & 0 & {\footnotesize \textcircled{\tiny 1}} & 0 & 0 & 0& 0 & 0 & 0& {\footnotesize \textcircled{\tiny 1}} & 0 & 0 & 0 \\
0 & 0 & 0& 0 & \kappa \delta t  & 0& 0 & 0 & 0 &  \kappa \delta t & 0& 0 & 0 & 0   \\
0 & 0 & 0& 0& 0 & {\footnotesize \textcircled{\tiny 2}} & 0 & 0 & {\footnotesize \textcircled{\tiny 2}} & 0& 0& 0 & 0 & 0 \\
0 & 0 & 0& 0& 0 &0  & \kappa \delta t &  \kappa \delta t & 0 & 0 & 0& 0& 0 & 0  \\
0 & 0 & 0& 0& 0 &0  & \kappa \delta t & \kappa \delta t & 0 & 0 & 0& 0& 0 & 0 \\
0 & 0 & 0& 0& 0 & {\footnotesize \textcircled{\tiny 2}}  & 0& 0 & {\footnotesize \textcircled{\tiny 2}} & 0& 0& 0 & 0 & 0   \\
0 & 0 & 0& 0 & \kappa \delta t  & 0& 0 & 0 & 0 &  \kappa \delta t & 0& 0 & 0 & 0   \\
0 & 0 & 0 & {\footnotesize \textcircled{\tiny 1}} & 0 & 0 & 0& 0 & 0 & 0& {\footnotesize \textcircled{\tiny 1}} & 0 & 0 & 0   \\
0 & 0 &  \kappa \delta t & 0& 0 & 0& 0 & 0 & 0& 0 & 0 &  \kappa \delta t & 0 & 0   \\
0 &  \kappa \delta t & 0 & 0& 0 & 0 & 0& 0 & 0 & 0 & 0 & 0 & \kappa \delta t & 0   \\
1 -4 \kappa \delta t & 0 & 0 & 0 & 0 & 0 & 0& 0 & 0 & 0 & 0 &
0 & 0 & 1 -4 \kappa \delta t \end{smallmatrix}\right),}
\end{eqnarray}
where \textcircled{\emph{n}} denotes $n$ diagonal zeros. Now, because of the form of the density matrix at $t=\delta t$, we
use the following ansatz for the density matrix for all times
\begin{eqnarray}\label{mat1}
\varepsilon(\rho_{4\mbox{\tiny GHZ}}) = { \left(
\begin{smallmatrix}
a & 0 & 0 & 0 & 0 & 0 & 0 & 0 & 0 & 0 & 0 & 0 & 0 & 0 & 0 & a   \\
0 & b & 0 & 0 & 0 & 0 & 0 & 0 & 0 & 0 & 0 & 0 & 0 & 0 & b & 0   \\
0 & 0 & b & 0 & 0 & 0 & 0 & 0 & 0 & 0 & 0 & 0 & 0 & b & 0 & 0   \\
0 & 0 & 0 & c & 0 & 0 & 0 & 0 & 0 & 0 & 0 & 0 & c & 0 & 0 & 0   \\
0 & 0 & 0 & 0 & b & 0 & 0 & 0 & 0 & 0 & 0 & b & 0 & 0 & 0 & 0   \\
0 & 0 & 0 & 0 & 0 & c & 0 & 0 & 0 & 0 & c & 0 & 0 & 0 & 0 & 0   \\
0 & 0 & 0 & 0 & 0 & 0 & c & 0 & 0 & c & 0 & 0 & 0 & 0 & 0 & 0   \\
0 & 0 & 0 & 0 & 0 & 0 & 0 & b & b & 0 & 0 & 0 & 0 & 0 & 0 & 0   \\
0 & 0 & 0 & 0 & 0 & 0 & 0 & b & b & 0 & 0 & 0 & 0 & 0 & 0 & 0   \\
0 & 0 & 0 & 0 & 0 & 0 & c & 0 & 0 & c & 0 & 0 & 0 & 0 & 0 & 0   \\
0 & 0 & 0 & 0 & 0 & c & 0 & 0 & 0 & 0 & c & 0 & 0 & 0 & 0 & 0   \\
0 & 0 & 0 & 0 & b & 0 & 0 & 0 & 0 & 0 & 0 & b & 0 & 0 & 0 & 0   \\
0 & 0 & 0 & c & 0 & 0 & 0 & 0 & 0 & 0 & 0 & 0 & c & 0 & 0 & 0   \\
0 & 0 & b & 0 & 0 & 0 & 0 & 0 & 0 & 0 & 0 & 0 & 0 & b & 0 & 0   \\
0 & b & 0 & 0 & 0 & 0 & 0 & 0 & 0 & 0 & 0 & 0 & 0 & 0 & b & 0   \\
a & 0 & 0 & 0 & 0 & 0 & 0 & 0 & 0 & 0 & 0 & 0 & 0 & 0 & 0 & a
\end{smallmatrix} \right).}
\end{eqnarray}
Inserting this matrix in the Lindblad equation, Eq.~({\ref{Lindblad}}),
gives us a set of three coupled differential equations
\begin{eqnarray}\label{diff1}
\left\{
\begin{array}{l}
\dot{a}(t) = 4k\Big(b(t)-a(t)\Big),\\
\dot{b}(t) = k\Big(a(t)-4b(t)+3c(t)\Big),\\
\dot{c}(t)=4k\Big(b(t)-c(t)\Big),
\end{array}\right.
\end{eqnarray}
subject to the initial conditions $a(0)=1/2$ and $b(0)=c(0)=0$ (see
Eq.~(\ref{ro0})). The solutions are readily given by
\begin{eqnarray}
\label{elements} \left\{
\begin{array}{l}
a(t) =\frac{1}{16}\left( 1 + 6 e^{-4 \kappa t} +  e^{-8 \kappa t}\right),\\
b(t) =\frac{1}{16}\left( 1 - e^{-8 \kappa t}\right),\\
c(t) =\frac{1}{16}\left( 1 - 2 e^{-4 \kappa t} + e^{-8 \kappa
t}\right).
\end{array}\right.
\end{eqnarray}
In fact, the infinitesimal temporal behavior of the density matrix
helped us to properly suggest the solution and consequently reduced
136 coupled differential equations to three coupled differential
equations which are readily solved. It is now easy to check that
$\varepsilon(\rho_{4\mbox{\tiny GHZ}})$, Eq.~(\ref{mat1}), exactly
satisfies the Lindblad equation, Eq.~({\ref{Lindblad}}), and the validity
of the ansatz is verfied.

Having $\varepsilon(\rho_{4\mbox{\tiny GHZ}})$ and $U_{\mbox{\tiny
tel}}$ which can be read off from Fig.~\ref{fig2}, it is
straightforward to compute $\rho_{out}$. Thus, the fidelity reads
\begin{eqnarray}
\label{xfidelity} & &F(\theta, \phi) = \frac{1}{2} \left[(1 + \sin^2
\theta \cos^2 \phi) + e^{-4 \kappa t} (\cos^2 \theta + \sin^2 \theta
\sin^2 \phi) \right],
\end{eqnarray}
and the average fidelity is given by
\begin{eqnarray}
\label{xfbar} \overline{F} = \frac{2}{3} + \frac{1}{3} e^{-4 \kappa
t}.
\end{eqnarray}

Now consider $(L_{2,y},L_{3,y},L_{4,y},L_{5,y})$ and assume
$\kappa_{2,y}=\kappa_{3,y}=\kappa_{4,y}=\kappa_{5,y}=\kappa$.
Similar to the previous case, using the infinitesimal time evolution
of the density matrix
\begin{eqnarray}
\varepsilon(\rho_{4\mbox{\tiny GHZ}})\Big|_{t=\delta t} =
\frac{1}{2}{\left(
\begin{smallmatrix}
1 -4 \kappa \delta t & 0 & 0 & 0 & 0& 0 & 0 & 0& 0 & 0 & 0 & 0 & 0 & 1-4 \kappa \delta t   \\
0 & \kappa \delta t & 0 & 0 & 0& 0 & 0 & 0 & 0 & 0 & 0 & 0 &  -\kappa \delta t & 0   \\
0 & 0 & \kappa \delta t & 0& 0 & 0& 0 & 0 & 0& 0 & 0 & -\kappa \delta t & 0 & 0   \\
0 & 0 & 0 & {\footnotesize \textcircled{\tiny 1}} & 0 & 0 & 0& 0 & 0 & 0& {\footnotesize \textcircled{\tiny 1}} & 0 & 0 & 0 \\
0 & 0 & 0& 0 & \kappa \delta t  & 0& 0 & 0 & 0&  -\kappa \delta t & 0& 0 & 0 & 0   \\
0 & 0 & 0& 0& 0 & {\footnotesize \textcircled{\tiny 2}} & 0 & 0 & {\footnotesize \textcircled{\tiny 2}}& 0& 0 & 0 & 0 & 0 \\
0 & 0 & 0& 0& 0 &0  & \kappa \delta t &  -\kappa \delta t & 0 & 0& 0& 0 & 0 & 0   \\
0 & 0 & 0& 0& 0 &0  & -\kappa \delta t & \kappa \delta t & 0 & 0& 0& 0 & 0 & 0   \\
0 & 0 & 0& 0& 0 & {\footnotesize \textcircled{\tiny 2}} & 0 & 0 & {\footnotesize \textcircled{\tiny 2}} & 0& 0& 0 & 0 & 0 \\
0 & 0 & 0& 0 & -\kappa \delta t  & 0& 0 & 0 & 0& \kappa \delta t & 0& 0 & 0 & 0   \\
0 & 0 & 0 & {\footnotesize \textcircled{\tiny 1}}& 0 & 0 & 0& 0 & 0 & 0& {\footnotesize \textcircled{\tiny 1}} & 0 & 0 & 0 \\
0 & 0 &  -\kappa \delta t & 0& 0 & 0& 0 & 0 & 0& 0 & 0 &  \kappa \delta t & 0 & 0 \\
0 &  -\kappa \delta t & 0 & 0& 0 & 0 & 0& 0 & 0 & 0 & 0 & 0 & \kappa \delta t & 0 \\
1 -4 \kappa \delta t & 0 & 0 & 0 & 0 & 0 & 0& 0 & 0 & 0 & 0
&0 & 0 & 1 -4 \kappa \delta t
\end{smallmatrix}\right),}
\end{eqnarray}
we take the following ansatz
\begin{eqnarray}
 \varepsilon(\rho_{4\mbox{\tiny GHZ}}) =
 \left(
\begin{smallmatrix}
a & 0 & 0 & 0 & 0 & 0 & 0& 0 & 0 & 0& 0 & 0 & 0 & 0 & 0 & a   \\
0 & b& 0 & 0& 0 & 0 & 0& 0 & 0 & 0 & 0 & 0 & 0 & 0 & -b & 0   \\
0 & 0 & b & 0& 0 & 0& 0 & 0 & 0& 0 & 0 & 0 & 0 & -b & 0 & 0   \\
0 & 0 & 0 & c & 0 & 0& 0 & 0 & 0& 0 & 0 & 0& c & 0 & 0 & 0   \\
0 & 0 & 0& 0 & b  & 0& 0 & 0 & 0& 0 & 0 & -b & 0& 0 & 0 & 0   \\
0 & 0 & 0& 0& 0 & c  & 0& 0 & 0 & 0 & c & 0& 0& 0 & 0 & 0   \\
0 & 0 & 0& 0& 0&0 & c  & 0& 0 & c & 0& 0& 0 & 0& 0 & 0   \\
0 & 0 & 0& 0& 0 &0  & 0& b & -b & 0 & 0 & 0& 0& 0 & 0 & 0   \\
0 & 0 & 0& 0& 0 &0  & 0& -b & b & 0 & 0 & 0& 0& 0 & 0 & 0   \\
0 & 0 & 0& 0& 0&0 & c  & 0& 0 & c & 0& 0& 0 & 0& 0 & 0   \\
0 & 0 & 0& 0& 0 & c  & 0& 0 & 0 & 0 & c & 0& 0& 0 & 0 & 0   \\
0 & 0 & 0& 0 &- b  & 0& 0 & 0 & 0& 0 & 0 & b & 0& 0 & 0 & 0   \\
0 & 0 & 0 & c & 0 & 0& 0 & 0 & 0& 0 & 0 & 0& c & 0 & 0 & 0   \\
0 & 0 &- b & 0& 0 & 0& 0 & 0 & 0& 0 & 0 & 0 & 0 & b & 0 & 0   \\
0 & -b & 0 & 0& 0 & 0 & 0& 0 & 0 & 0 & 0 & 0 & 0 & 0 & b & 0   \\
a & 0 & 0 & 0 & 0 & 0 & 0& 0 & 0 & 0& 0 & 0 & 0 & 0 & 0 & a
\end{smallmatrix}
\right).
\end{eqnarray}
Inserting this matrix in the Lindblad equation, Eq.({\ref{Lindblad}}),
gives the previous set of coupled differential equations, Eq.~(\ref{diff1}), and consequently the solutions agree with
Eq.~(\ref{elements}). For this case the fidelity becomes
\begin{eqnarray}
\label{yfidelity} & &F(\theta, \phi) = \frac{1}{2} \left[1 +( \sin^2
\theta \sin^2 \phi + \cos^2 \theta) e^{-4 \kappa t} + \sin^2 \theta
\cos^2 \phi e^{-8 \kappa t} \right],
\\   \nonumber
\end{eqnarray}
and the average fidelity reads
\begin{eqnarray}
\label{yfbar} \overline{F} = \frac{1}{2} +\frac{1}{3}  e^{-4 \kappa
t} +\frac{1}{6} e^{-8 \kappa t}.
\end{eqnarray}

For the third case consider $(L_{2,z},L_{3,z},L_{4,z},L_{5,z})$ and
assume $\kappa_{2,z}=\kappa_{3,z}=\kappa_{4,z}=\kappa_{5,z}=\kappa$.
The infinitesimal time evolution of the density matrix gives
\begin{eqnarray}
\varepsilon(\rho_{4\mbox{\tiny GHZ}})\Big|_{t=\delta t} =
\frac{1}{2} \left( |0\rangle^{\otimes 4} \langle 0 |^{\otimes 4} +
|1\rangle ^{\otimes 4}\langle 1| ^{\otimes 4}\right) +
\frac{1-8\kappa \delta t}{2} \left(|0\rangle ^{\otimes 4} \langle 1|
^{\otimes 4}+ |1\rangle ^{\otimes 4}\langle 0 |^{\otimes 4} \right).
\end{eqnarray}
So the ansatz is
\begin{eqnarray}
\varepsilon(\rho_{4\mbox{\tiny GHZ}}) = a \left( |0\rangle^{\otimes
4} \langle 0 |^{\otimes 4} + |1\rangle ^{\otimes 4}\langle 1|
^{\otimes 4}\right) + b\left(|0\rangle ^{\otimes 4} \langle 1|
^{\otimes 4}+ |1\rangle ^{\otimes 4}\langle 0 |^{\otimes 4} \right).
\end{eqnarray}
Inserting this matrix in the Lindblad equation, Eq.~({\ref{Lindblad}}),
results in
\begin{eqnarray}\label{diff2}
\left\{
\begin{array}{l}
\dot{a}(t) = 0,\\
\dot{b}(t) =-8k\, b(t) ,
\end{array}\right.
\end{eqnarray}
subject to the initial condition $a(0)=b(0)=1/2$. The solution is
\begin{equation}
\label{zmatrix} \varepsilon(\rho_{4\mbox{\tiny GHZ}}) = \frac{1}{2}
\left( |0\rangle^{\otimes 4} \langle 0 |^{\otimes 4} + |1\rangle
^{\otimes 4}\langle 1| ^{\otimes 4}\right) + \frac{1}{2} e^{-8
\kappa t} \left(|0\rangle ^{\otimes 4} \langle 1| ^{\otimes 4}+
|1\rangle ^{\otimes 4}\langle 0 |^{\otimes 4} \right).
\end{equation}
Also, the fidelity and its average read
\begin{eqnarray}
\label{z,f,fbar}
\begin{array}{l}\displaystyle
F(\theta, \phi) = 1 - \frac{1}{2} \left(1 - e^{-8 \kappa t} \right)
\sin^2 \theta, \\\\ \displaystyle\overline{F} = \frac{2}{3} +
\frac{1}{3} e^{-8 \kappa t}.
\end{array}
\end{eqnarray}

The next noisy channel is the isotropic noisy channel. For this case, the
master equation involves twelve Lindblad operators
$(L_{2,\alpha},L_{3,\alpha},L_{4,\alpha},L_{5,\alpha})$ with
$\alpha\in\{x,y,z\}$. At $t=\delta t$ we have
\begin{eqnarray}
\varepsilon(\rho_{4\mbox{\tiny GHZ}})\Big|_{t=\delta t} =
\frac{1}{2}{\left(
\begin{smallmatrix}
1 -8 \kappa \delta t & 0 & 0 & 0 & 0 & 0 & 0 & 0& 0 & 0 & 0 & 0 & 0 & 1-16 \kappa \delta t \\
0 & 2\kappa \delta t & 0 & 0& 0 & 0 & 0& 0 & 0 & 0 & 0 & 0 & 0 & 0 \\
0 & 0 & 2\kappa \delta t & 0& 0 & 0& 0 & 0 & 0 & 0 & 0 & 0 & 0 & 0 \\
0 & 0 & 0 & {\footnotesize \textcircled{\tiny 1}} & 0 & 0 & 0& 0 & 0 & 0& {\footnotesize \textcircled{\tiny 1}} & 0& 0 & 0 \\
0 & 0 & 0& 0 & 2\kappa \delta t  & 0& 0 & 0 & 0& 0 & 0 &  0 & 0& 0 \\
0 & 0 & 0& 0& 0 &{\footnotesize \textcircled{\tiny 2}} & 0& 0 & {\footnotesize \textcircled{\tiny 2}} &0 & 0 & 0& 0& 0 \\
0 & 0 & 0& 0& 0 &0 & 2\kappa \delta t & 0 & 0 & 0& 0& 0 & 0 & 0  \\
0 & 0 & 0& 0& 0 &0 & 0 & 2\kappa \delta t & 0 & 0 & 0& 0& 0 & 0 \\
0 & 0 & 0& 0& 0 & {\footnotesize \textcircled{\tiny 2}} & 0 & 0 & {\footnotesize \textcircled{\tiny 2}} & 0& 0& 0 & 0 & 0   \\
0 & 0 & 0& 0 & 0  & 0& 0 & 0 & 0&  2\kappa \delta t & 0& 0 & 0 & 0   \\
0 & 0 & 0 & {\footnotesize \textcircled{\tiny 1}} & 0 & 0& 0 & 0 & 0& 0 & {\footnotesize \textcircled{\tiny 1}} & 0 & 0 & 0   \\
0 & 0 &  0 & 0& 0 & 0& 0 & 0 & 0& 0 & 0 &  2\kappa \delta t & 0 & 0   \\
0 &  0 & 0 & 0& 0 & 0 & 0& 0 & 0 & 0 & 0 & 0 & 2\kappa \delta t & 0   \\
1 -16 \kappa \delta t & 0 & 0 & 0 & 0& 0 & 0 & 0& 0 & 0 & 0
&0 & 0 & 1 -8 \kappa \delta t
\end{smallmatrix}\right).}
\end{eqnarray}
So we take the ansatz
\begin{eqnarray}
\varepsilon(\rho_{4\mbox{\tiny GHZ}}) =
 \left(
\begin{smallmatrix}
a & 0 & 0 & 0 & 0 & 0 & 0& 0 & 0 & 0& 0 & 0 & 0 & 0 & 0 & d   \\
0 & b& 0 & 0& 0 & 0 & 0& 0 & 0 & 0 & 0 & 0 & 0 & 0 & 0 & 0   \\
0 & 0 & b & 0& 0 & 0& 0 & 0 & 0& 0 & 0 & 0 & 0 & 0 & 0 & 0   \\
0 & 0 & 0 & c & 0 & 0& 0 & 0 & 0& 0 & 0 & 0& 0 & 0 & 0 & 0   \\
0 & 0 & 0& 0 & b  & 0& 0 & 0 & 0& 0 & 0 & 0 & 0& 0 & 0 & 0   \\
0 & 0 & 0& 0& 0 & c  & 0& 0 & 0 & 0 & 0 & 0& 0& 0 & 0 & 0   \\
0 & 0 & 0& 0& 0&0 & c  & 0& 0 & 0 & 0& 0& 0 & 0& 0 & 0   \\
0 & 0 & 0& 0& 0 &0  & 0& b & 0 & 0 & 0 & 0& 0& 0 & 0 & 0   \\
0 & 0 & 0& 0& 0 &0  & 0& 0 & b & 0 & 0 & 0& 0& 0 & 0 & 0   \\
0 & 0 & 0& 0& 0&0 & 0  & 0& 0 & c & 0& 0& 0 & 0& 0 & 0   \\
0 & 0 & 0& 0& 0 & 0  & 0& 0 & 0 & 0 & c & 0& 0& 0 & 0 & 0   \\
0 & 0 & 0& 0 & 0 & 0& 0 & 0 & 0& 0 & 0 & b & 0& 0 & 0 & 0   \\
0 & 0 & 0 & 0 & 0 & 0& 0 & 0 & 0& 0 & 0 & 0& c & 0 & 0 & 0   \\
0 & 0 &0 & 0& 0 & 0& 0 & 0 & 0& 0 & 0 & 0 & 0 & b & 0 & 0   \\
0 & 0 & 0 & 0& 0 & 0 & 0& 0 & 0 & 0 & 0 & 0 & 0 & 0 & b & 0   \\
d & 0 & 0 & 0 & 0 & 0 & 0& 0 & 0 & 0& 0 & 0 & 0 & 0 & 0 & a
\end{smallmatrix}
\right).
\end{eqnarray}
Inserting this solution in the Lindblad equation, Eq.~({\ref{Lindblad}}),
we find
\begin{eqnarray}
\left\{
\begin{array}{l}
\dot{a}(t) = 8k\Big(b(t)-a(t)\Big),\\
\dot{b}(t) = 2k\Big(a(t)-4b(t)+3c(t)\Big),\\
\dot{c}(t)=8k\Big(b(t)-c(t)\Big),\\
\dot{d}(t)=-16k\,d(t),\\
\end{array}\right.
\end{eqnarray}
subject to the initial conditions $a(0)=d(0)=1/2$ and $b(0)=c(0)=0$.
The solutions are
\begin{eqnarray}\left\{
\begin{array}{l}
a(t) =\frac{1}{16}\Big( 1 + 6 e^{-8 \kappa t} + e^{-16 \kappa t}\Big),\\
b(t) =\frac{1}{16}\Big( 1 - e^{-16 \kappa t}\Big),\\
c(t) = \frac{1}{16}\Big(1 - 2 e^{-8 \kappa t} +  e^{-16 \kappa t}\Big),\\
d(t) = \frac{1}{2} e^{-16 \kappa t}.
\end{array}\right.
\end{eqnarray}
Also the fidelity is
\begin{eqnarray}
& &F(\theta, \phi) = \frac{1}{2} \left[ 1 + e^{-8 \kappa t} \cos^2
\theta + e^{-16 \kappa t}  \sin^2 \theta \right],
\end{eqnarray}
and
\begin{eqnarray}
\label{dfbar} \overline{F} = \frac{1}{6} \left( 3 +  e^{-8 \kappa t}
+ 2 e^{-16 \kappa t} \right).
\end{eqnarray}

To this end, we only considered the noisy channels with the same
axis. Now, as a different-axis noisy channel, consider
$(L_{2,x},L_{3,y},L_{4,z},L_{5,x})$ noise with
$\kappa_{2,x}=\kappa_{3,y}=\kappa_{4,z}=\kappa_{5,x}=\kappa$ that
exhibits the effects of noises in different directions. After an
infinitesimal time interval and using the Lindblad equation, the
density matrix  can be written as
\begin{eqnarray}
\varepsilon(\rho_{4\mbox{\tiny GHZ}})\Big|_{t=\delta t} =
\frac{1}{2}{\left(
\begin{smallmatrix}
1 -6 \kappa \delta t & 0 & 0 & 0 & 0 & 0& 0 & 0 & 0 & 0 & 0 & 1-10 \kappa \delta t \\
0 & \kappa \delta t & 0 & 0& 0 & 0 & 0 & 0 & 0 & 0 & \kappa \delta t & 0 \\
0 & 0 & {\footnotesize \textcircled{\tiny 2}} & 0& 0 & 0 & 0 & 0 & 0 & {\footnotesize \textcircled{\tiny 2}} & 0 & 0 \\
0 & 0 & 0 & \kappa \delta t & 0 & 0 &0 & 0 & -\kappa \delta t & 0& 0 & 0 \\
0 & 0 & 0 & 0& {\footnotesize \textcircled{\tiny 2}}& 0 & 0 & {\footnotesize \textcircled{\tiny 2}}& 0 & 0 & 0 & 0 \\
0 & 0 & 0& 0 & 0& \kappa \delta t  & \kappa \delta t & 0 & 0 & 0& 0 & 0 \\
0 & 0 & 0& 0 & 0& \kappa \delta t  & \kappa \delta t & 0 & 0 & 0& 0 & 0 \\
0 & 0 & 0 & 0& {\footnotesize \textcircled{\tiny 2}}& 0 & 0 & {\footnotesize \textcircled{\tiny 2}} & 0 & 0 & 0 & 0 \\
0 & 0 & 0 & -\kappa \delta t & 0 & 0& 0& 0 & \kappa \delta t & 0& 0 & 0 \\
0 & 0 &{\footnotesize \textcircled{\tiny 2}}& 0 & 0 & 0 & 0& 0 & 0 & {\footnotesize \textcircled{\tiny 2}} & 0 & 0 \\
0 & \kappa \delta t & 0 & 0& 0 & 0 & 0& 0 & 0 & 0 & \kappa \delta t & 0 \\
1 -10 \kappa \delta t & 0 & 0 & 0 & 0& 0 & 0 & 0 & 0
&0 & 0 & 1 -6 \kappa \delta t
\end{smallmatrix}\right).}
\end{eqnarray}
So, the elements of the density matrix for all time can be read off
as
\begin{eqnarray}
\varepsilon(\rho_{4\mbox{\tiny GHZ}}) =
 \left(
\begin{smallmatrix}
a & 0 & 0 & 0 & 0 & 0 & 0& 0 & 0 & 0& 0 & 0 & 0 & 0 & 0 & g   \\
0 & b& 0 & 0& 0 & 0 & 0& 0 & 0 & 0 & 0 & 0 & 0 & 0 & h & 0   \\
0 & 0 & c & 0& 0 & 0& 0 & 0 & 0& 0 & 0 & 0 & 0 & m & 0 & 0   \\
0 & 0 & 0 & d & 0 & 0& 0 & 0 & 0& 0 & 0 & 0& k & 0 & 0 & 0   \\
0 & 0 & 0& 0 & b  & 0& 0 & 0 & 0& 0 & 0 & n & 0& 0 & 0 & 0   \\
0 & 0 & 0& 0& 0 & d  & 0& 0 & 0 & 0 & k & 0& 0& 0 & 0 & 0   \\
0 & 0 & 0& 0& 0&0 & d  & 0& 0 & f & 0& 0& 0 & 0& 0 & 0   \\
0 & 0 & 0& 0& 0 &0  & 0& b & h & 0 & 0 & 0& 0& 0 & 0 & 0   \\
0 & 0 & 0& 0& 0 &0  & 0& h & b & 0 & 0 & 0& 0& 0 & 0 & 0   \\
0 & 0 & 0& 0& 0&0 & f  & 0& 0 & d & 0& 0& 0 & 0& 0 & 0   \\
0 & 0 & 0& 0& 0 & k  & 0& 0 & 0 & 0 & d & 0& 0& 0 & 0 & 0   \\
0 & 0 & 0& 0 & n & 0& 0 & 0 & 0& 0 & 0 & b & 0& 0 & 0 & 0   \\
0 & 0 & 0 & k & 0 & 0& 0 & 0 & 0& 0 & 0 & 0& d & 0 & 0 & 0   \\
0 & 0 & m & 0& 0 & 0& 0 & 0 & 0& 0 & 0 & 0 & 0 & c & 0 & 0   \\
0 & h & 0 & 0& 0 & 0 & 0& 0 & 0 & 0 & 0 & 0 & 0 & 0 & b & 0   \\
g & 0 & 0 & 0 & 0 & 0 & 0& 0 & 0 & 0& 0 & 0 & 0 & 0 & 0 & a
\end{smallmatrix}
\right),
\end{eqnarray}
which leads to two sets of four and six coupled differential
equations, namely
\begin{eqnarray}
\left\{
\begin{array}{l}
\dot{a}(t) = 3\kappa\Big(b(t)-a(t)\Big),\\
\dot{b}(t) = \kappa\Big(a(t)-3b(t)+2d(t)\Big),\\
\dot{c}(t) = 3\kappa\Big(d(t)-c(t)\Big),\\
\dot{d}(t) = \kappa\Big(2b(t)-3d(t)+c(t)\Big),
\end{array}\right.
\end{eqnarray}
and
\begin{eqnarray}
\left\{
\begin{array}{l}
\dot{f}(t) = \kappa\Big(-5f(t)+2h(t)-m(t)\Big),\\
\dot{g}(t) = \kappa\Big(-5g(t)+2h(t)-n(t)\Big),\\
\dot{h}(t) = \kappa\Big(f(t)+g(t)-5h(t)-k(t)\Big),\\
\dot{k}(t) = \kappa\Big(-h(t)-5k(t)+m(t)+n(t)\Big),\\
\dot{m}(t)=\kappa\Big(-f(t)+2k(t)-5m(t)\Big),\\
\dot{n}(t)=\kappa\Big(-g(t)+2k(t)-5n(t)\Big),
\end{array}\right.
\end{eqnarray}
subject to $a(0)=g(0)=1/2$ and
$b(0)=c(0)=d(0)=f(0)=h(0)=k(0)=m(0)=n(0)=0$. The solutions are
readily found
\begin{eqnarray}\left\{
\begin{array}{l}
a(t) =e^{2 \kappa t}g(t)=\frac{1}{16}\Big( 1 + 3 e^{-2 \kappa t} +3 e^{-4 \kappa t} + e^{-6 \kappa t}\Big),\\
b(t) =e^{2 \kappa t}h(t)=-e^{2 \kappa t}n(t)=\frac{1}{16}\Big( 1 + e^{-2 \kappa t} - e^{-4 \kappa t} - e^{-6 \kappa t}\Big),\\
c(t) =-e^{2 \kappa t}m(t)=\frac{1}{16}\Big( 1 - 3 e^{-2 \kappa t} +3 e^{-4 \kappa t} - e^{-6 \kappa t}\Big),\\
d(t) =e^{2 \kappa t}f(t)=-e^{2 \kappa t}k(t)=\frac{1}{16}\Big( 1 -
e^{-2 \kappa t} - e^{-4 \kappa t} + e^{-6 \kappa t}\Big).
\end{array}\right.
\end{eqnarray}
Thus, the fidelity, $F(\theta, \phi)$, and its average, $\overline{F}$, are given by
\begin{eqnarray}
& &F(\theta, \phi) = \frac{1}{2} \left[ 1 + e^{-2 \kappa t} \cos^2
\theta + e^{-4 \kappa t} \sin^2
\theta \cos^2 \phi + e^{-6 \kappa t}  \sin^2 \theta \sin^2 \phi \right],
\end{eqnarray}
and
\begin{eqnarray}
\label{mixfbar} \overline{F} = \frac{1}{6} \left( 3 + e^{-2 \kappa t}+ e^{-4 \kappa t}
+ e^{-6 \kappa t} \right).
\end{eqnarray}

\begin{table}
\caption{\label{tab1}Summary of $F(\theta, \phi)$ and $\overline{F}$
through various noisy channels.}
\begin{ruledtabular}
\begin{tabular}{c|ccc}
{} & Noise & 3GHZ & 4GHZ \\ \hline
{} & Pauli-X & $\frac{1}{2} \bigg[(1 + \sin^2 \theta \cos^2 \phi$) & $\frac{1}{2} \bigg[ (1 + \sin^2 \theta \cos^2 \phi$) \\
{} & {} & $+ e^{-4 \kappa t} (\cos^2 \theta + \sin^2 \theta \sin^2 \phi) \bigg]$ & $+ e^{-4 \kappa t} (\cos^2 \theta + \sin^2 \theta \sin^2 \phi) \bigg]$ \\
$F(\theta,\phi)$ & Pauli-Y &$\frac{1}{2} \bigg[1 + \sin^2 \theta \sin^2 \phi e^{-2 \kappa t} +\cos^2 \theta e^{-4 \kappa t}$   & $\frac{1}{2} \bigg[1 + (\sin^2 \theta \sin^2 \phi +\cos^2 \theta) e^{-4 \kappa t}$    \\
{} & {} & $+ \sin^2 \theta \cos^2 \phi e^{-6 \kappa t} \bigg]$ & $+\sin^2 \theta \cos^2 \phi e^{-8 \kappa t} \bigg]$\\
{} & Pauli-Z & $1 - \frac{1}{2} (1 -e^{-6 \kappa t}) \sin^2 \theta$  &   $1 - \frac{1}{2} (1 - e^{-8\kappa t})\sin^2 \theta$\\
{} & isotropic & $\frac{1}{2} (1 + \cos^2 \theta e^{-8 \kappa t} +\sin^2 \theta e^{-12 \kappa t} )$ & $\frac{1}{2} (1 + \cos^2 \theta e^{-8 \kappa t} + \sin^2 \theta e^{-16 \kappa t} )$\\\hline
{} & Pauli-X & $\frac{2}{3} +\frac{1}{3} e^{-4 \kappa t}$ & $\frac{2}{3} + \frac{1}{3} e^{-4\kappa t}$\\
$\overline{F}$ & Pauli-Y &$\frac{1}{6} (3 + e^{-2 \kappa t} + e^{-4 \kappa t} + e^{-6 \kappa t})$    & $\frac{1}{6} (3 + 2 e^{-4 \kappa t} + e^{-8 \kappa t})$\\
{} & Pauli-Z & $\frac{2}{3} +\frac{1}{3} e^{-6 \kappa t}$ & $\frac{2}{3} + \frac{1}{3} e^{-8\kappa t}$\\
{} & isotropic & $\frac{1}{6} (3 + e^{-8 \kappa t} + 2 e^{-12 \kappa t} )$ & $\frac{1}{6} (3 + e^{-8 \kappa t} + 2 e^{-16 \kappa t})$
\end{tabular}
\end{ruledtabular}
\end{table}

\begin{figure}
\centering
\includegraphics[width=8cm]{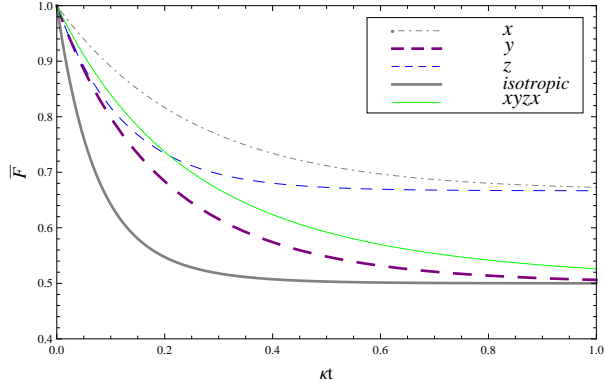}
\caption{\label{fig7} The plot of time dependence of average
fidelity through noisy channels for
4GHZ state.}
\end{figure}

\begin{figure}
\centerline{\begin{tabular}{ccc}
\includegraphics[width=8cm]{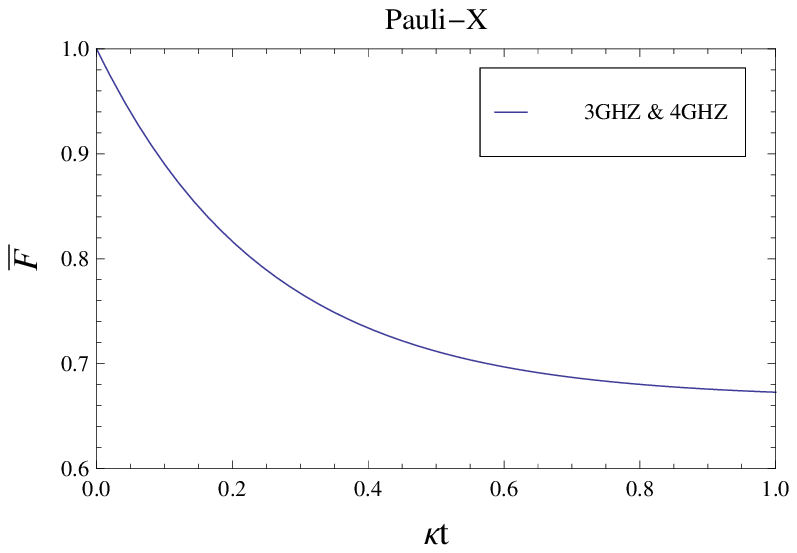}
 &\hspace{2.cm}&
\includegraphics[width=8cm]{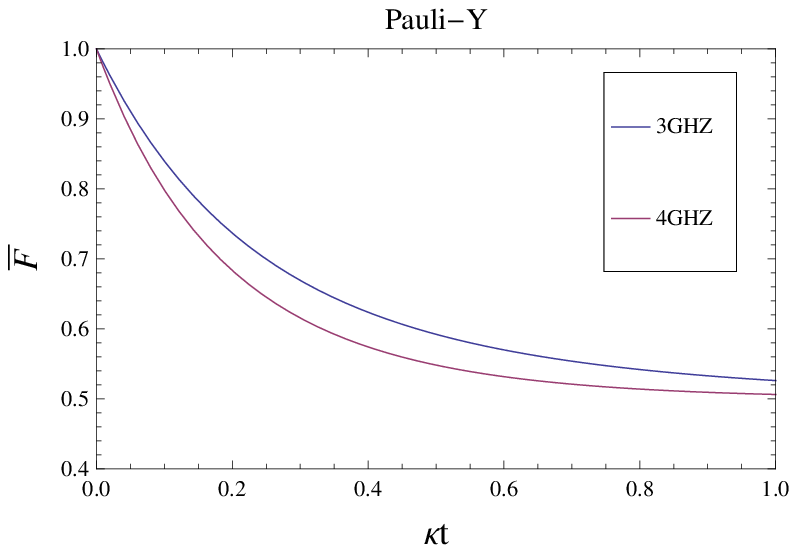}\\
\includegraphics[width=8cm]{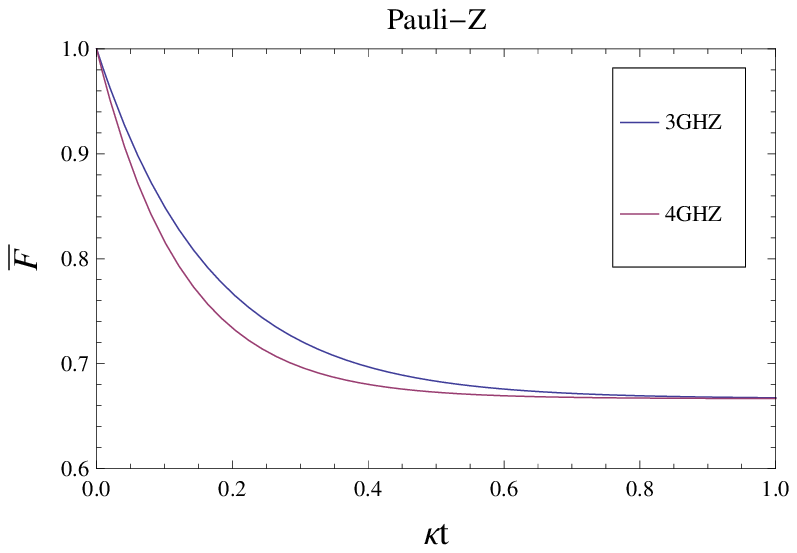}
 &\hspace{2.cm}&
\includegraphics[width=8cm]{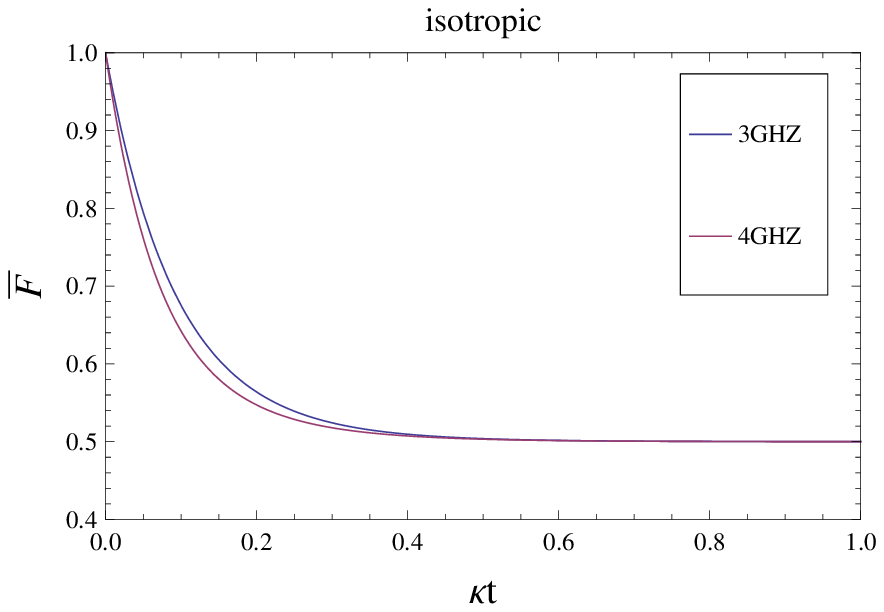}
\end{tabular}}
\caption{\label{fig3} The plot of time dependence of average
fidelity for Pauli-X (left up),
Pauli-Y (right up),
Pauli-Z (left down), and isotropic
(right down) noisy channels.}
\end{figure}

In Table \ref{tab1}, a summary of fidelity and average fidelity for
3GHZ \cite{jung08-2} and 4GHZ states is reported and compared. Also,
their average fidelity versus time is depicted in Fig.~\ref{fig3}
for various noisy channels. Comparing 3GHZ and 4GHZ states shows
that for $(L_{2,x},L_{3,x},L_{4,x},L_{5,x})$ noise both states have
the same fidelity. This result also agrees with Bell state
$|\beta_{00}\rangle=\frac{1}{\sqrt{2}}(|00\rangle+|11\rangle)$
\cite{0h02}. However, for other cases 3GHZ state is more robust,
i.e., loses less quantum information in the quantum teleportation
process with respect to 4GHZ state. Note that, for the isotropic
case, the fidelities are approximately equal. These results and
those obtained in Refs.~\cite{jung08-2,0h02} show that increasing
the number of qubits can enhance the rate of information lost in
quantum teleportation process. Moreover, using a proper ansatz for
the density matrix, we reduced the number of coupled differential
equations from 136 to at most four coupled equations.
Fig.~\ref{fig7} shows average fidelity for 4GHZ state through
various noises. As it can be seen from the figure,
$(L_{2,x},L_{3,x},L_{4,x},L_{5,x})$ noise does lose less quantum
information with respect to others. The next noise with small
information lost is $(L_{2,x},L_{3,y},L_{4,z},L_{5,x})$  for $\kappa
t<0.2$. However, for $\kappa t>0.2$,
$(L_{2,z},L_{3,z},L_{4,z},L_{5,z})$ noise represents a better
behavior. Moreover, the isotropic noise and the noise in $y$
direction always result in low fidelity quantum teleportation. In
the following sections, we exactly solve the Lindblad equation for
5GHZ and 6GHZ states through two types of noisy channels.

\section{Five-qubit GHZ state with noisy channels}\label{sec4}
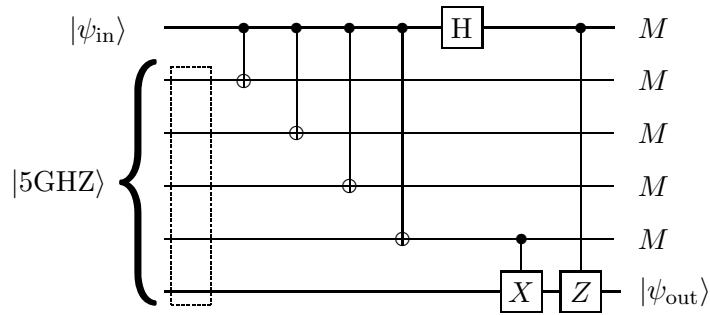
\begin{figure}[t]
%
%
\begin{picture}(260,120)
\put(0,114){\makebox(10,10){$|\psi_{\text{in}}\rangle$}}
\put(-15,56){\makebox(10,10){$|5\mbox{GHZ}\rangle$}}
\put(10,38){\resizebox{20pt}{70pt}{\{}}

\put(32.5,15){\dashbox{1}(15,90){}}

\put(30,120){\line(1,0){105}}\put(150,120){\line(1,0){50}}
\put(30,100){\line(1,0){170}}
\put(30,80){\line(1,0){170}}
\put(30,60){\line(1,0){170}}
\put(30,40){\line(1,0){170}}
\put(30,20){\line(1,0){127.5}}\put(172.5,20){\line(1,0){7.5}}\put(195,20){\line(1,0){7.5}}

\put(60,120){\circle*{4}}
\put(60,120){\line(0,-1){22.5}}
\put(60,100){\circle{5}}

\put(80,120){\circle*{4}}
\put(80,120){\line(0,-1){42.5}}
\put(80,80){\circle{5}}

\put(100,120){\circle*{4}}
\put(100,120){\line(0,-1){62.5}}
\put(100,60){\circle{5}}

\put(120,120){\circle*{4}}
\put(120,120){\line(0,-1){82.5}}
\put(120,40){\circle{5}}

\put(135,112.5){\framebox(15,15){H}}

\put(165,40){\circle*{4}}
\put(165,40){\line(0,-1){12.5}}
\put(157.5,12.5){\framebox(15,15){$X$}}

\put(187.5,120){\circle*{4}}
\put(187.5,120){\line(0,-1){92.5}}
\put(180,12.5){\framebox(15,15){$Z$}}

\put(210,115){\makebox(10,10){$M$}}
\put(210,95){\makebox(10,10){$M$}}
\put(210,75){\makebox(10,10){$M$}}
\put(210,55){\makebox(10,10){$M$}}
\put(210,35){\makebox(10,10){$M$}}
\put(220,15){\makebox(10,10){$|\psi_{\text{out}}\rangle$ }}
\end{picture}
\caption{\label{fig4}
   A circuit for quantum teleportation through noisy channels with 5GHZ state.
   The five top lines belong to Alice and the bottom line to Bob.
   $M$ denotes measurement and the dotted box represents noisy channel.
   The Lindblad operator is turned on inside the dotted box.}
\end{figure}

In this section, we teleport 5GHZ state through noisy channels as
depicted in Fig.~\ref{fig4}. For this case the solution of the
Lindblad equation is a $32\times32$ matrix that results in a set of
32 diagonal and 496 off-diagonal coupled differential equations.
However, we show that the number of required equations can be
considerably reduced by choosing appropriate ansatz for the density
matrix.

First, consider ($L_{2,x}$,$L_{3,x}$,$L_{4,x}$,$L_{5,x}$,$L_{6,x}$)
noise and assume
$\kappa_{2,x}=\kappa_{3,x}=\kappa_{4,x}=\kappa_{5,x}=\kappa_{6,x}=\kappa$.
The infinitesimal time evolution of the density matrix now reads

\begin{eqnarray}
\hspace{-1cm} \varepsilon(\rho_{5\mbox{\tiny GHZ}})\Big|_{t=\delta
t} = \frac{1}{2}{ \left(
\begin{smallmatrix}
1- 5 \kappa \delta t & 0 & 0 & 0 & 0 & 0 & 0 & 0 & 0 & 0 & 0 & 0 & 0 & 0 & 0 & 0 & 0 & 1- 5 \kappa \delta t  \\
0 & \kappa \delta t & 0 & 0 & 0 & 0 & 0 & 0& 0 & 0 & 0 & 0 & 0 & 0 & 0 &  0 & \kappa \delta t & 0   \\
0 & 0 & \kappa \delta t & 0 & 0 & 0 & 0 & 0 & 0 & 0 & 0 & 0 & 0 & 0 & 0 & \kappa \delta t & 0 & 0   \\
0 & 0 & 0 & {\footnotesize \textcircled{\tiny 1}} & 0 & 0 & 0 & 0 & 0 & 0 & 0 & 0 & 0 & 0 & {\footnotesize \textcircled{\tiny 1}} & 0 & 0 & 0   \\
0 & 0 & 0& 0 & \kappa \delta t  & 0 & 0& 0 & 0 & 0 & 0 & 0 & 0 & \kappa \delta t & 0& 0 & 0 & 0   \\
0 & 0 & 0& 0 &0 & {\footnotesize \textcircled{\tiny 3}} & 0 & 0& 0 & 0 & 0 & 0 & 0 & {\footnotesize \textcircled{\tiny 3}} & 0& 0 & 0 & 0   \\
0 & 0 & 0& 0 &0 & 0 & \kappa \delta t  & 0 & 0 & 0 & 0 & \kappa \delta t & 0 & 0 & 0& 0 & 0 & 0   \\
0 & 0 & 0& 0 &0 &0 & 0& {\footnotesize \textcircled{\tiny 6}} & 0 & 0 & {\footnotesize \textcircled{\tiny 6}} & 0 & 0 & 0 & 0& 0 & 0 & 0   \\
0& 0 & 0 & 0& 0& 0 & 0 & 0 & \kappa \delta t & \kappa \delta t & 0 & 0 & 0& 0& 0 & 0 & 0 & 0 \\
0& 0 & 0 & 0& 0& 0 & 0 & 0 & \kappa \delta t & \kappa \delta t & 0 & 0 & 0& 0& 0 & 0 & 0 & 0 \\
0 & 0 & 0& 0 &0 &0 & 0& {\footnotesize \textcircled{\tiny 6}} & 0& 0 & {\footnotesize \textcircled{\tiny 6}} & 0 & 0 & 0 & 0& 0 & 0 & 0   \\
0 & 0 & 0& 0 &0 & 0 & \kappa \delta t  & 0 & 0 & 0 & 0 & \kappa \delta t & 0& 0 & 0& 0 & 0 & 0   \\
0 & 0 & 0& 0 &0 & {\footnotesize \textcircled{\tiny 3}} & 0 & 0 & 0 & 0 & 0 & 0 & {\footnotesize \textcircled{\tiny 3}} &0 & 0& 0 & 0 & 0   \\
0 & 0 & 0& 0 & \kappa \delta t & 0 & 0 & 0 & 0 & 0 & 0 &0 & 0 & \kappa \delta t & 0& 0 & 0 & 0   \\
0 & 0 & 0 & {\footnotesize \textcircled{\tiny 1}} & 0 & 0 & 0 & 0 & 0 & 0 & 0 & 0 & 0 & 0 & {\footnotesize \textcircled{\tiny 1}} & 0 & 0 & 0   \\
0 & 0 & \kappa \delta t & 0 & 0 & 0 & 0 & 0 & 0& 0& 0 & 0 & 0 & 0 & 0 & \kappa \delta t & 0 & 0   \\
0 & \kappa \delta t & 0 & 0 & 0 & 0 & 0 & 0& 0 & 0 & 0& 0 & 0 & 0 & 0 & 0 & \kappa \delta t & 0   \\
1- 5 \kappa \delta t & 0 & 0 & 0 & 0 & 0 &
0 & 0 & 0 & 0 & 0 & 0 & 0 & 0& 0 & 0 & 0
& 1- 5 \kappa \delta t
\end{smallmatrix} \right).}
\hspace{0.5cm}
\end{eqnarray}
So we take the ansatz as
\begin{eqnarray}
\hspace{-1cm} \varepsilon(\rho_{5\mbox{\tiny GHZ}}) = \left(
\begin{smallmatrix}
a & 0 & 0 & 0 & 0 & 0 & 0 & 0 & 0 & 0 & 0 & 0 & 0 & 0 & 0 & 0 & 0 & a \\
0 & b & 0 & 0 & 0 & 0 & 0 & 0 & 0 & 0 & 0 & 0 & 0 & 0 & 0 &  0 & b & 0   \\
0 & 0 & b & 0 & 0 & 0 & 0 & 0 & 0 & 0 & 0 & 0 & 0 & 0 & 0 & b & 0 & 0   \\
0 & 0 & 0 & \tiny\textcircled{\emph{c}}_{1} & 0 & 0 & 0 & 0 & 0 & 0 & 0 & 0 & 0 & 0 & \tiny\textcircled{\emph{c}}_{1} & 0 & 0 & 0   \\
0 & 0 & 0& 0 & b & 0 & 0& 0 & 0 & 0 & 0 &0 & 0 & b & 0& 0 & 0 & 0   \\
0 & 0 & 0& 0 &0 & \tiny\textcircled{\emph{c}}_{3} & 0 & 0 & 0 & 0 & 0 & 0 & \tiny\textcircled{\emph{c}}_{3} &0 & 0& 0 & 0 & 0   \\
0 & 0 & 0& 0 &0 & 0 & b  & 0 & 0 & 0 & 0 & b & 0 & 0 & 0& 0 & 0 & 0   \\
0 & 0 & 0& 0 &0 &0 & 0& \tiny\textcircled{\emph{c}}_{6} & 0 & 0 & \tiny\textcircled{\emph{c}}_{6} &0 &  0 & 0 & 0& 0 & 0 & 0   \\
0& 0 & 0 & 0& 0& 0 & 0 & 0& b & b & 0 & 0 & 0& 0& 0 & 0 & 0 & 0 \\
0& 0 & 0 & 0& 0& 0 & 0 & 0 & b & b & 0 & 0 & 0& 0& 0 & 0 & 0 & 0 \\
0 & 0 & 0& 0 &0 &0 & 0& \tiny\textcircled{\emph{c}}_{6} & 0 & 0 & \tiny\textcircled{\emph{c}}_{6} &0 &  0 & 0 & 0& 0 & 0 & 0   \\
0 & 0 & 0& 0 &0 & 0 & b & 0 & 0 & 0 & 0 & b & 0& 0 & 0& 0 & 0 & 0   \\
0 & 0 & 0& 0 &0 & \tiny\textcircled{\emph{c}}_{3} & 0 & 0 & 0 & 0 & 0 & 0 & \tiny\textcircled{\emph{c}}_{3} &0 & 0& 0 & 0 & 0 \\
0 & 0 & 0& 0 & b &0 & 0& 0 & 0 & 0 & 0 &0 & 0 & b & 0& 0 & 0 & 0   \\
0 & 0 & 0 &\tiny\textcircled{\emph{c}}_{1} & 0 & 0& 0 & 0 & 0& 0 & 0 & 0 & 0 & 0 & \tiny\textcircled{\emph{c}}_{1} & 0 & 0 & 0   \\
0 & 0 & b & 0 & 0 & 0 & 0& 0 & 0 & 0& 0 & 0 & 0 & 0 & 0 & b & 0 & 0   \\
0 & b& 0 & 0 & 0& 0 & 0 & 0& 0 & 0 & 0& 0 & 0 & 0 & 0 &  0 & b & 0   \\
a & 0 & 0 & 0 & 0 & 0 & 0& 0 & 0 & 0& 0 & 0 & 0 & 0 & 0 & 0 & 0 & a
\end{smallmatrix} \right).
\hspace{0.5cm}
\end{eqnarray}
Here $\textcircled{\emph{c}}_{n}$ denotes
$n$ diagonal $c$.

Now inserting this matrix in Lindblad equation, Eq.~(\ref{Lindblad}), four
coupled differential equations are obtained as follows
\begin{eqnarray}
\left\{
\begin{array}{l}
\dot{a}(t) = 5k\Big(b(t)-a(t)\Big),\\
\dot{b}(t) = k\Big(a(t)-5b(t)+4c(t)\Big),\\
\dot{c}(t)=2k\Big(b(t)-c(t)\Big).\\
\end{array}\right.
\end{eqnarray}
Solving this set of equations with the initial conditions
$a(0)=1/2$, $b(0)=c(0)=0$, leads to the following solution
\begin{eqnarray}
\left\{
\begin{array}{l}
a(t) =\frac{1}{32}\Big( 1 + 10 e^{-4 \kappa t} + 5e^{-8 \kappa t}\Big),\\
b(t) =\frac{1}{32}\Big( 1 + 2 e^{-4 \kappa t}- 3e^{-8 \kappa t}\Big),\\
c(t) = \frac{1}{32}\Big(1 - 2 e^{-4 \kappa t} +  e^{-8 \kappa
t}\Big).
\end{array}\right.
\end{eqnarray}
Substituting $\varepsilon(\rho_{5\mbox{\tiny GHZ}})$ in
Eq.~(\ref{out1}) and using Eqs.~(\ref{fidelity}) and (\ref{average})
fidelity and its average are given by
\begin{eqnarray}
F(\theta, \phi) &=& \frac{1}{2} \left[ 1 + \sin^2 \theta \cos^2
\phi+ e^{-4 \kappa t} (\cos^2 \theta +
     \sin^2 \theta \sin^2 \phi )\right],\\
\overline{F} &=& \frac{1}{3} \left( 2 + e^ {-4 \kappa t} \right).
\end{eqnarray}

For ($L_{2,z}$,$L_{3,z}$,$L_{4,z}$,$L_{5,z}$,$L_{6,z}$) noise with
$\kappa_{2,z}=\kappa_{3,z}=\kappa_{4,z}=\kappa_{5,z}=\kappa_{6,z}=\kappa$,
the infinitesimal evolution matrix is
\begin{eqnarray}
\varepsilon(\rho_{5\mbox{\tiny GHZ}})\Big|_{t=\delta t} =
\frac{1}{2} \left( |0\rangle^{\otimes 5} \langle 0 |^{\otimes 5} +
|1\rangle ^{\otimes 5}\langle 1| ^{\otimes 5}\right) +
\frac{1-10\kappa \delta t}{2} \left(|0\rangle ^{\otimes 5} \langle
1| ^{\otimes 5}+ |1\rangle ^{\otimes 5}\langle 0 |^{\otimes 5}
\right).
\end{eqnarray}
Using the ansatz
\begin{eqnarray}
 \varepsilon(\rho_{5\mbox{\tiny GHZ}}) =
a \left( |0\rangle^{\otimes 5} \langle 0 |^{\otimes 5} + |1\rangle
^{\otimes 5}\langle 1| ^{\otimes 5}\right) + b\left(|0\rangle
^{\otimes 5} \langle 1| ^{\otimes 5}+ |1\rangle ^{\otimes 5}\langle
0 |^{\otimes 5} \right),
\end{eqnarray}
we obtain two coupled equations
\begin{eqnarray}
\left\{
\begin{array}{l}
\dot{a}(t) =0,\\
\dot{b}(t) = -10kb(t),\\
\end{array}\right.
\end{eqnarray}
subject to $a(0)=1/2$, $b(0)=0$. Therefore, the density matrix reads
\begin{equation}
\label{zmatrix5} \varepsilon(\rho_{5\mbox{\tiny GHZ}}) = \frac{1}{2}
\left( |0\rangle^{\otimes 5} \langle 0 |^{\otimes 5} + |1\rangle
^{\otimes 5}\langle 1| ^{\otimes 5}\right) + \frac{1}{2} e^{-10
\kappa t} \left(|0\rangle ^{\otimes 5} \langle 1| ^{\otimes 5}+
|1\rangle ^{\otimes 5}\langle 0 |^{\otimes 5} \right),
\end{equation}
and the fidelity and its average are given by
\begin{eqnarray}
F(\theta, \phi) &=&  1 - \frac{1}{2} \left( 1 - e^{-10 \kappa t}\right) \sin^2 \theta ,\\
\overline{F} &=& \frac{1}{3} \left( 2 + e^ {-10 \kappa t} \right).
\end{eqnarray}

\section{Six-qubit GHZ state with noisy channels}\label{sec5}
A quantum circuit for  teleportation through noisy channels with
6GHZ state is depicted in Fig.~\ref{fig5}. In the dotted box the
Lindblad operators act on the $64\times64$ density matrix that
involves five Alice's qubits and one Bob's qubits. The Lindblad
equation, Eq.~(\ref{Lindblad}), leads to 64 diagonal and 2016 off-diagonal
linear coupled differential equations. However, similar to previous
sections, we first study infinitesimal temporal behavior of the
density matrix and use a proper ansatz to considerably reduce the
number of required equations.

\begin{figure}[t]
%
%
\begin{picture}(320,140)
\put(5,134){\makebox(10,10){$|\psi_{\text{in}}\rangle$}}
\put(-20,64){\makebox(10,10){$|6\mbox{GHZ}\rangle$}}
\put(5,42){\resizebox{20pt}{85pt}{\{}}

\put(32.5,15){\dashbox{1}(17.5,110){}}

\put(30,140){\line(1,0){135}}\put(180,140){\line(1,0){60}}
\put(30,120){\line(1,0){210}}
\put(30,100){\line(1,0){210}}
\put(30,80){\line(1,0){210}}
\put(30,60){\line(1,0){210}}
\put(30,40){\line(1,0){210}}
\put(30,20){\line(1,0){160}}\put(205,20){\line(1,0){12.5}}\put(232.5,20){\line(1,0){7.5}}

\put(65,140){\circle*{4}}
\put(65,140){\line(0,-1){22.5}}
\put(65,120){\circle{5}}

\put(85,140){\circle*{4}}
\put(85,140){\line(0,-1){42.5}}
\put(85,100){\circle{5}}

\put(105,140){\circle*{4}}
\put(105,140){\line(0,-1){62.5}}
\put(105,80){\circle{5}}

\put(125,140){\circle*{4}}
\put(125,140){\line(0,-1){82.5}}
\put(125,60){\circle{5}}

\put(145,140){\circle*{4}}
\put(145,140){\line(0,-1){102.5}}
\put(145,40){\circle{5}}

\put(165,132.5){\framebox(15,15){H}}

\put(197.5,40){\circle*{4}}
\put(197.5,40){\line(0,-1){12.5}}
\put(190,12.5){\framebox(15,15){$X$}}

\put(225,140){\circle*{4}}
\put(225,140){\line(0,-1){112.5}}
\put(217.5,12.5){\framebox(15,15){$Z$}}

\put(242.5,135){\makebox(10,10){$M$}}
\put(242.5,115){\makebox(10,10){$M$}}
\put(242.5,95){\makebox(10,10){$M$}}
\put(242.5,75){\makebox(10,10){$M$}}
\put(242.5,55){\makebox(10,10){$M$}}
\put(242.5,35){\makebox(10,10){$M$}}
\put(252.5,15){\makebox(10,10){$|\psi_{\text{out}}\rangle$ }}
\end{picture}
\caption{\label{fig5}
   A circuit for quantum teleportation through noisy channels with 6GHZ state.
   The six top lines belong to Alice and the bottom line to Bob.
   $M$ denotes measurement and the dotted box represents noisy channel.
   The Lindblad operator is turned on inside the dotted box.}
\end{figure}

For ($L_{2,x}$,$L_{3,x}$,$L_{4,x}$,$L_{5,x}$,$L_{6,x},L_{7,x}$)
noise and
$\kappa_{2,x}=\kappa_{3,x}=\kappa_{4,x}=\kappa_{5,x}=\kappa_{6,x}=\kappa_{7,x}=\kappa$,
the Lindblad operators after an infinitesimal time transform the
input density matrix
$\rho(0)=|6\mbox{GHZ}\rangle\langle6\mbox{GHZ}|$ to
\begin{eqnarray}
\hspace{-1cm}\varepsilon(\rho_{6\mbox{\tiny GHZ}})\Big|_{t=\delta
t}= \frac {1}{2}\left(\mbox{\footnotesize$
\begin{smallmatrix}
     1 - 6 \kappa \delta t  & 0 & 0 & 0 & 0 & 0 & 0 & 0 & 0 & 0 & 0 & 0 & 0 & 0 & 0 & 0 & 0 & 0 & 0 & 0 & 0 &  1 - 6 \kappa \delta t \\
    0 & \kappa \delta t & 0 & 0 & 0 & 0 & 0 & 0 & 0 & 0 & 0 & 0 & 0 & 0 & 0 & 0 & 0 & 0 & 0 & 0 & \kappa \delta t & 0 \\
    0 & 0 & \kappa \delta t& 0 & 0 & 0 & 0 & 0 & 0 & 0 & 0 & 0 & 0 & 0 & 0 & 0 & 0 & 0 & 0 & \kappa \delta t & 0 & 0 \\
    0 & 0 & 0 & {\footnotesize \textcircled{\tiny 1}} & 0 & 0 & 0 & 0 & 0 & 0 & 0 & 0 & 0 & 0 & 0 & 0 & 0 & 0 & {\footnotesize \textcircled{\tiny 1}} & 0 & 0 & 0 \\
    0 & 0 & 0 & 0 & \kappa \delta t & 0 & 0 & 0 & 0 & 0 & 0 & 0 & 0 & 0 & 0 & 0 & 0 & \kappa \delta t & 0 & 0 & 0 & 0 \\
    0 & 0 & 0 & 0 & 0 & {\footnotesize \textcircled{\tiny 3}} & 0 & 0 & 0 & 0 & 0 & 0 & 0 & 0 & 0 & 0 & {\footnotesize \textcircled{\tiny 3}} & 0 & 0 & 0 & 0 & 0 \\
    0 & 0 & 0 & 0 & 0 & 0 & \kappa \delta t & 0 & 0 & 0 & 0 & 0 & 0 & 0 & 0 & \kappa \delta t & 0 & 0 & 0 & 0 & 0 & 0 \\
    0 & 0 & 0 & 0 & 0 & 0 & 0 & {\footnotesize \textcircled{\tiny 7}} & 0 & 0 & 0 & 0 & 0 & 0 & {\footnotesize \textcircled{\tiny 7}} & 0 & 0 & 0 & 0 & 0 & 0 & 0 \\
    0 & 0 & 0 & 0 & 0 & 0 & 0 & 0 & \kappa \delta t & 0 & 0 & 0 & 0 & \kappa \delta t & 0 & 0 & 0 & 0 & 0 & 0 & 0 & 0 \\
    0 & 0 & 0 & 0 & 0 & 0 & 0 & 0 & 0 & {\footnotesize \textcircled{\tiny 14}} & 0 & 0 & {\footnotesize \textcircled{\tiny 14}} & 0 & 0 & 0 & 0 & 0 & 0 & 0 & 0 & 0 \\
    0 & 0 & 0 & 0 & 0 & 0 & 0 & 0 & 0 & 0 & \kappa \delta t & \kappa \delta t & 0 & 0 & 0 & 0 & 0 & 0 & 0 & 0 & 0 & 0 \\
    0 & 0 & 0 & 0 & 0 & 0 & 0 & 0 & 0 & 0 & \kappa \delta t & \kappa \delta t & 0 & 0 & 0 & 0 & 0 & 0 & 0 & 0 & 0 & 0 \\
    0 & 0 & 0 & 0 & 0 & 0 & 0 & 0 & 0 & {\footnotesize \textcircled{\tiny 14}} & 0 & 0 &{\footnotesize \textcircled{\tiny 14}} & 0 & 0 & 0 & 0 & 0 & 0 & 0 & 0 & 0 \\
    0 & 0 & 0 & 0 & 0 & 0 & 0 & 0 & \kappa \delta t & 0 & 0 & 0 & 0 & \kappa \delta t & 0 & 0 & 0 & 0 & 0 & 0 & 0 & 0 \\
    0 & 0 & 0 & 0 & 0 & 0 & 0 & {\footnotesize \textcircled{\tiny 7}} & 0 & 0 & 0 & 0 & 0 & 0 & {\footnotesize \textcircled{\tiny 7}} & 0 & 0 & 0 & 0 & 0 & 0 & 0 \\
    0 & 0 & 0 & 0 & 0 & 0 & \kappa \delta t & 0 & 0 & 0 & 0 & 0 & 0 & 0 & 0 & \kappa \delta t & 0 & 0 & 0 & 0 & 0 & 0 \\
    0 & 0 & 0 & 0 & 0 & {\footnotesize \textcircled{\tiny 3}} & 0 & 0 & 0 & 0 & 0 & 0 & 0 & 0 & 0 & 0 & {\footnotesize \textcircled{\tiny 3}} & 0 & 0 & 0 & 0 & 0 \\
    0 & 0 & 0 & 0 & \kappa \delta t & 0 & 0 & 0 & 0 & 0 & 0 & 0 & 0 & 0 & 0 & 0 & 0 & \kappa \delta t & 0 & 0 & 0 & 0 \\
    0 & 0 & 0 & {\footnotesize \textcircled{\tiny 1}} & 0 & 0 & 0 & 0 & 0 & 0 & 0 & 0 & 0 & 0 & 0 & 0 & 0 & 0 & {\footnotesize \textcircled{\tiny 1}} & 0 & 0 & 0 \\
    0 & 0 & \kappa \delta t& 0 & 0 & 0 & 0 & 0 & 0 & 0 & 0 & 0 & 0 & 0 & 0 & 0 & 0 & 0 & 0 & \kappa \delta t & 0 & 0 \\
    0 & \kappa \delta t & 0 & 0 & 0 & 0 & 0 & 0 & 0 & 0 & 0 & 0 & 0 & 0 & 0 & 0 & 0 & 0 & 0 & 0 & \kappa \delta t & 0 \\
     1 - 6 \kappa \delta t  & 0 & 0 & 0 & 0 & 0 & 0 & 0 & 0 & 0 & 0 & 0 & 0 & 0 & 0 & 0 & 0 & 0 & 0 & 0 & 0 &  1 - 6 \kappa \delta t
\end{smallmatrix}$}\right).
\end{eqnarray}
So consider
the ansatz
\begin{eqnarray}
\hspace{-2.3cm}\varepsilon(\rho_{6\mbox{\tiny GHZ}})=
\left(\mbox{\footnotesize$
\begin{smallmatrix}
a & 0 & 0 & 0 & 0 & 0 & 0 & 0 & 0 & 0 & 0 & 0 & 0 & 0 & 0 & 0 & 0 & 0 & 0 & 0 & 0 & 0 & 0 & 0 & 0 & 0 & 0 & 0 & 0 & 0 & 0 & 0 & 0 & 0 & 0 & 0 & 0 & 0 & 0 & 0 & 0 & 0 & 0 & 0 & 0 & 0 & 0 & a \\
0 & b & 0 & 0 & 0 & 0 & 0 & 0 & 0 & 0 & 0 & 0 & 0 & 0 & 0 & 0 & 0 & 0 & 0 & 0 & 0 & 0 & 0 & 0 & 0 & 0 & 0 & 0 & 0 & 0 & 0 & 0 & 0 & 0 & 0 & 0 & 0 & 0 & 0 & 0 & 0 & 0 & 0 & 0 & 0 & 0 & b & 0 \\
0 & 0 & b & 0 & 0 & 0 & 0 & 0 & 0 & 0 & 0 & 0 & 0 & 0 & 0 & 0 & 0 & 0 & 0 & 0 & 0 & 0 & 0 & 0 & 0 & 0 & 0 & 0 & 0 & 0 & 0 & 0 & 0 & 0 & 0 & 0 & 0 & 0 & 0 & 0 & 0 & 0 & 0 & 0 & 0 & b & 0 & 0 \\
0 & 0 & 0 & c & 0 & 0 & 0 & 0 & 0 & 0 & 0 & 0 & 0 & 0 & 0 & 0 & 0 & 0 & 0 & 0 & 0 & 0 & 0 & 0 & 0 & 0 & 0 & 0 & 0 & 0 & 0 & 0 & 0 & 0 & 0 & 0 & 0 & 0 & 0 & 0 & 0 & 0 & 0 & 0 & c & 0 & 0 & 0 \\
0 & 0 & 0 & 0 & b & 0 & 0 & 0 & 0 & 0 & 0 & 0 & 0 & 0 & 0 & 0 & 0 & 0 & 0 & 0 & 0 & 0 & 0 & 0 & 0 & 0 & 0 & 0 & 0 & 0 & 0 & 0 & 0 & 0 & 0 & 0 & 0 & 0 & 0 & 0 & 0 & 0 & 0 & b & 0 & 0 & 0 & 0 \\
0 & 0 & 0 & 0 & 0 & \tiny\textcircled{\emph{c}} & 0 & 0 & 0 & 0 & 0 & 0 & 0 & 0 & 0 & 0 & 0 & 0 & 0 & 0 & 0 & 0 & 0 & 0 & 0 & 0 & 0 & 0 & 0 & 0 & 0 & 0 & 0 & 0 & 0 & 0 & 0 & 0 & 0 & 0 & 0 & 0 & \tiny\textcircled{\emph{c}} & 0 & 0 & 0 & 0 & 0 \\
0 & 0 & 0 & 0 & 0 & 0 & d & 0 & 0 & 0 & 0 & 0 & 0 & 0 & 0 & 0 & 0 & 0 & 0 & 0 & 0 & 0 & 0 & 0 & 0 & 0 & 0 & 0 & 0 & 0 & 0 & 0 & 0 & 0 & 0 & 0 & 0 & 0 & 0 & 0 & 0 & d & 0 & 0 & 0 & 0 & 0 & 0 \\
0 & 0 & 0 & 0 & 0 & 0 & 0 & b & 0 & 0 & 0 & 0 & 0 & 0 & 0 & 0 & 0 & 0 & 0 & 0 & 0 & 0 & 0 & 0 & 0 & 0 & 0 & 0 & 0 & 0 & 0 & 0 & 0 & 0 & 0 & 0 & 0 & 0 & 0 & 0 & b & 0 & 0 & 0 & 0 & 0 & 0 & 0 \\
0 & 0 & 0 & 0 & 0 & 0 & 0 & 0 & \tiny\textcircled{\emph{c}} & 0 & 0 & 0 & 0 & 0 & 0 & 0 & 0 & 0 & 0 & 0 & 0 & 0 & 0 & 0 & 0 & 0 & 0 & 0 & 0 & 0 & 0 & 0 & 0 & 0 & 0 & 0 & 0 & 0 & 0 & \tiny\textcircled{\emph{c}} & 0 & 0 & 0 & 0 & 0 & 0 & 0 & 0 \\
0 & 0 & 0 & 0 & 0 & 0 & 0 & 0 & 0 & d & 0 & 0 & 0 & 0 & 0 & 0 & 0 & 0 & 0 & 0 & 0 & 0 & 0 & 0 & 0 & 0 & 0 & 0 & 0 & 0 & 0 & 0 & 0 & 0 & 0 & 0 & 0 & 0 & d & 0 & 0 & 0 & 0 & 0 & 0 & 0 & 0 & 0 \\
0 & 0 & 0 & 0 & 0 & 0 & 0 & 0 & 0 & 0 & c & 0 & 0 & 0 & 0 & 0 & 0 & 0 & 0 & 0 & 0 & 0 & 0 & 0 & 0 & 0 & 0 & 0 & 0 & 0 & 0 & 0 & 0 & 0 & 0 & 0 & 0 & c & 0 & 0 & 0 & 0 & 0 & 0 & 0 & 0 & 0 & 0 \\
0 & 0 & 0 & 0 & 0 & 0 & 0 & 0 & 0 & 0 & 0 & \tiny\textcircled{\emph{d}} & 0 & 0 & 0 & 0 & 0 & 0 & 0 & 0 & 0 & 0 & 0 & 0 & 0 & 0 & 0 & 0 & 0 & 0 & 0 & 0 & 0 & 0 & 0 & 0 & \tiny\textcircled{\emph{d}} & 0 & 0 & 0 & 0 & 0 & 0 & 0 & 0 & 0 & 0 & 0 \\
0 & 0 & 0 & 0 & 0 & 0 & 0 & 0 & 0 & 0 & 0 & 0 & c & 0 & 0 & 0 & 0 & 0 & 0 & 0 & 0 & 0 & 0 & 0 & 0 & 0 & 0 & 0 & 0 & 0 & 0 & 0 & 0 & 0 & 0 & c & 0 & 0 & 0 & 0 & 0 & 0 & 0 & 0 & 0 & 0 & 0 & 0 \\
0 & 0 & 0 & 0 & 0 & 0 & 0 & 0 & 0 & 0 & 0 & 0 & 0 & b & 0 & 0 & 0 & 0 & 0 & 0 & 0 & 0 & 0 & 0 & 0 & 0 & 0 & 0 & 0 & 0 & 0 & 0 & 0 & 0 & b & 0 & 0 & 0 & 0 & 0 & 0 & 0 & 0 & 0 & 0 & 0 & 0 & 0 \\
0 & 0 & 0 & 0 & 0 & 0 & 0 & 0 & 0 & 0 & 0 & 0 & 0 & 0 & \tiny\textcircled{\emph{c}} & 0 & 0 & 0 & 0 & 0 & 0 & 0 & 0 & 0 & 0 & 0 & 0 & 0 & 0 & 0 & 0 & 0 & 0 & \tiny\textcircled{\emph{c}} & 0 & 0 & 0 & 0 & 0 & 0 & 0 & 0 & 0 & 0 & 0 & 0 & 0 & 0 \\
0 & 0 & 0 & 0 & 0 & 0 & 0 & 0 & 0 & 0 & 0 & 0 & 0 & 0 & 0 & d & 0 & 0 & 0 & 0 & 0 & 0 & 0 & 0 & 0 & 0 & 0 & 0 & 0 & 0 & 0 & 0 & d & 0 & 0 & 0 & 0 & 0 & 0 & 0 & 0 & 0 & 0 & 0 & 0 & 0 & 0 & 0 \\
0 & 0 & 0 & 0 & 0 & 0 & 0 & 0 & 0 & 0 & 0 & 0 & 0 & 0 & 0 & 0 & c & 0 & 0 & 0 & 0 & 0 & 0 & 0 & 0 & 0 & 0 & 0 & 0 & 0 & 0 & c & 0 & 0 & 0 & 0 & 0 & 0 & 0 & 0 & 0 & 0 & 0 & 0 & 0 & 0 & 0 & 0 \\
0 & 0 & 0 & 0 & 0 & 0 & 0 & 0 & 0 & 0 & 0 & 0 & 0 & 0 & 0 & 0 & 0 & \tiny\textcircled{\emph{d}} & 0 & 0 & 0 & 0 & 0 & 0 & 0 & 0 & 0 & 0 & 0 & 0 & \tiny\textcircled{\emph{d}} & 0 & 0 & 0 & 0 & 0 & 0 & 0 & 0 & 0 & 0 & 0 & 0 & 0 & 0 & 0 & 0 & 0 \\
0 & 0 & 0 & 0 & 0 & 0 & 0 & 0 & 0 & 0 & 0 & 0 & 0 & 0 & 0 & 0 & 0 & 0 & \tiny\textcircled{\emph{c}} & 0 & 0 & 0 & 0 & 0 & 0 & 0 & 0 & 0 & 0 & \tiny\textcircled{\emph{c}} & 0 & 0 & 0 & 0 & 0 & 0 & 0 & 0 & 0 & 0 & 0 & 0 & 0 & 0 & 0 & 0 & 0 & 0 \\
0 & 0 & 0 & 0 & 0 & 0 & 0 & 0 & 0 & 0 & 0 & 0 & 0 & 0 & 0 & 0 & 0 & 0 & 0 & \tiny\textcircled{\emph{d}} & 0 & 0 & 0 & 0 & 0 & 0 & 0 & 0 & \tiny\textcircled{\emph{d}} & 0 & 0 & 0 & 0 & 0 & 0 & 0 & 0 & 0 & 0 & 0 & 0 & 0 & 0 & 0 & 0 & 0 & 0 & 0 \\
0 & 0 & 0 & 0 & 0 & 0 & 0 & 0 & 0 & 0 & 0 & 0 & 0 & 0 & 0 & 0 & 0 & 0 & 0 & 0 & c & 0 & 0 & 0 & 0 & 0 & 0 & c & 0 & 0 & 0 & 0 & 0 & 0 & 0 & 0 & 0 & 0 & 0 & 0 & 0 & 0 & 0 & 0 & 0 & 0 & 0 & 0 \\
0 & 0 & 0 & 0 & 0 & 0 & 0 & 0 & 0 & 0 & 0 & 0 & 0 & 0 & 0 & 0 & 0 & 0 & 0 & 0 & 0 & d & 0 & 0 & 0 & 0 & d & 0 & 0 & 0 & 0 & 0 & 0 & 0 & 0 & 0 & 0 & 0 & 0 & 0 & 0 & 0 & 0 & 0 & 0 & 0 & 0 & 0 \\
0 & 0 & 0 & 0 & 0 & 0 & 0 & 0 & 0 & 0 & 0 & 0 & 0 & 0 & 0 & 0 & 0 & 0 & 0 & 0 & 0 & 0 & \tiny\textcircled{\emph{c}} & 0 & 0 & \tiny\textcircled{\emph{c}} & 0 & 0 & 0 & 0 & 0 & 0 & 0 & 0 & 0 & 0 & 0 & 0 & 0 & 0 & 0 & 0 & 0 & 0 & 0 & 0 & 0 & 0 \\
0 & 0 & 0 & 0 & 0 & 0 & 0 & 0 & 0 & 0 & 0 & 0 & 0 & 0 & 0 & 0 & 0 & 0 & 0 & 0 & 0 & 0 & 0 & b & b & 0 & 0 & 0 & 0 & 0 & 0 & 0 & 0 & 0 & 0 & 0 & 0 & 0 & 0 & 0 & 0 & 0 & 0 & 0 & 0 & 0 & 0 & 0 \\
0 & 0 & 0 & 0 & 0 & 0 & 0 & 0 & 0 & 0 & 0 & 0 & 0 & 0 & 0 & 0 & 0 & 0 & 0 & 0 & 0 & 0 & 0 & b & b & 0 & 0 & 0 & 0 & 0 & 0 & 0 & 0 & 0 & 0 & 0 & 0 & 0 & 0 & 0 & 0 & 0 & 0 & 0 & 0 & 0 & 0 & 0 \\
0 & 0 & 0 & 0 & 0 & 0 & 0 & 0 & 0 & 0 & 0 & 0 & 0 & 0 & 0 & 0 & 0 & 0 & 0 & 0 & 0 & 0 & \tiny\textcircled{\emph{c}} & 0 & 0 & \tiny\textcircled{\emph{c}} & 0 & 0 & 0 & 0 & 0 & 0 & 0 & 0 & 0 & 0 & 0 & 0 & 0 & 0 & 0 & 0 & 0 & 0 & 0 & 0 & 0 & 0 \\
0 & 0 & 0 & 0 & 0 & 0 & 0 & 0 & 0 & 0 & 0 & 0 & 0 & 0 & 0 & 0 & 0 & 0 & 0 & 0 & 0 & d & 0 & 0 & 0 & 0 & d & 0 & 0 & 0 & 0 & 0 & 0 & 0 & 0 & 0 & 0 & 0 & 0 & 0 & 0 & 0 & 0 & 0 & 0 & 0 & 0 & 0 \\
0 & 0 & 0 & 0 & 0 & 0 & 0 & 0 & 0 & 0 & 0 & 0 & 0 & 0 & 0 & 0 & 0 & 0 & 0 & 0 & c & 0 & 0 & 0 & 0 & 0 & 0 & c & 0 & 0 & 0 & 0 & 0 & 0 & 0 & 0 & 0 & 0 & 0 & 0 & 0 & 0 & 0 & 0 & 0 & 0 & 0 & 0 \\
0 & 0 & 0 & 0 & 0 & 0 & 0 & 0 & 0 & 0 & 0 & 0 & 0 & 0 & 0 & 0 & 0 & 0 & 0 & \tiny\textcircled{\emph{d}} & 0 & 0 & 0 & 0 & 0 & 0 & 0 & 0 & \tiny\textcircled{\emph{d}} & 0 & 0 & 0 & 0 & 0 & 0 & 0 & 0 & 0 & 0 & 0 & 0 & 0 & 0 & 0 & 0 & 0 & 0 & 0 \\
0 & 0 & 0 & 0 & 0 & 0 & 0 & 0 & 0 & 0 & 0 & 0 & 0 & 0 & 0 & 0 & 0 & 0 & \tiny\textcircled{\emph{c}} & 0 & 0 & 0 & 0 & 0 & 0 & 0 & 0 & 0 & 0 & \tiny\textcircled{\emph{c}} & 0 & 0 & 0 & 0 & 0 & 0 & 0 & 0 & 0 & 0 & 0 & 0 & 0 & 0 & 0 & 0 & 0 & 0 \\
0 & 0 & 0 & 0 & 0 & 0 & 0 & 0 & 0 & 0 & 0 & 0 & 0 & 0 & 0 & 0 & 0 & \tiny\textcircled{\emph{d}} & 0 & 0 & 0 & 0 & 0 & 0 & 0 & 0 & 0 & 0 & 0 & 0 & \tiny\textcircled{\emph{d}} & 0 & 0 & 0 & 0 & 0 & 0 & 0 & 0 & 0 & 0 & 0 & 0 & 0 & 0 & 0 & 0 & 0 \\
0 & 0 & 0 & 0 & 0 & 0 & 0 & 0 & 0 & 0 & 0 & 0 & 0 & 0 & 0 & 0 & c & 0 & 0 & 0 & 0 & 0 & 0 & 0 & 0 & 0 & 0 & 0 & 0 & 0 & 0 & c & 0 & 0 & 0 & 0 & 0 & 0 & 0 & 0 & 0 & 0 & 0 & 0 & 0 & 0 & 0 & 0 \\
0 & 0 & 0 & 0 & 0 & 0 & 0 & 0 & 0 & 0 & 0 & 0 & 0 & 0 & 0 & d & 0 & 0 & 0 & 0 & 0 & 0 & 0 & 0 & 0 & 0 & 0 & 0 & 0 & 0 & 0 & 0 & d & 0 & 0 & 0 & 0 & 0 & 0 & 0 & 0 & 0 & 0 & 0 & 0 & 0 & 0 & 0 \\
0 & 0 & 0 & 0 & 0 & 0 & 0 & 0 & 0 & 0 & 0 & 0 & 0 & 0 & \tiny\textcircled{\emph{c}} & 0 & 0 & 0 & 0 & 0 & 0 & 0 & 0 & 0 & 0 & 0 & 0 & 0 & 0 & 0 & 0 & 0 & 0 & \tiny\textcircled{\emph{c}} & 0 & 0 & 0 & 0 & 0 & 0 & 0 & 0 & 0 & 0 & 0 & 0 & 0 & 0 \\
0 & 0 & 0 & 0 & 0 & 0 & 0 & 0 & 0 & 0 & 0 & 0 & 0 & b & 0 & 0 & 0 & 0 & 0 & 0 & 0 & 0 & 0 & 0 & 0 & 0 & 0 & 0 & 0 & 0 & 0 & 0 & 0 & 0 & b & 0 & 0 & 0 & 0 & 0 & 0 & 0 & 0 & 0 & 0 & 0 & 0 & 0 \\
0 & 0 & 0 & 0 & 0 & 0 & 0 & 0 & 0 & 0 & 0 & 0 & c & 0 & 0 & 0 & 0 & 0 & 0 & 0 & 0 & 0 & 0 & 0 & 0 & 0 & 0 & 0 & 0 & 0 & 0 & 0 & 0 & 0 & 0 & c & 0 & 0 & 0 & 0 & 0 & 0 & 0 & 0 & 0 & 0 & 0 & 0 \\
0 & 0 & 0 & 0 & 0 & 0 & 0 & 0 & 0 & 0 & 0 & \tiny\textcircled{\emph{d}} & 0 & 0 & 0 & 0 & 0 & 0 & 0 & 0 & 0 & 0 & 0 & 0 & 0 & 0 & 0 & 0 & 0 & 0 & 0 & 0 & 0 & 0 & 0 & 0 & \tiny\textcircled{\emph{d}} & 0 & 0 & 0 & 0 & 0 & 0 & 0 & 0 & 0 & 0 & 0 \\
0 & 0 & 0 & 0 & 0 & 0 & 0 & 0 & 0 & 0 & c & 0 & 0 & 0 & 0 & 0 & 0 & 0 & 0 & 0 & 0 & 0 & 0 & 0 & 0 & 0 & 0 & 0 & 0 & 0 & 0 & 0 & 0 & 0 & 0 & 0 & 0 & c & 0 & 0 & 0 & 0 & 0 & 0 & 0 & 0 & 0 & 0 \\
0 & 0 & 0 & 0 & 0 & 0 & 0 & 0 & 0 & d & 0 & 0 & 0 & 0 & 0 & 0 & 0 & 0 & 0 & 0 & 0 & 0 & 0 & 0 & 0 & 0 & 0 & 0 & 0 & 0 & 0 & 0 & 0 & 0 & 0 & 0 & 0 & 0 & d & 0 & 0 & 0 & 0 & 0 & 0 & 0 & 0 & 0 \\
0 & 0 & 0 & 0 & 0 & 0 & 0 & 0 & \tiny\textcircled{\emph{c}} & 0 & 0 & 0 & 0 & 0 & 0 & 0 & 0 & 0 & 0 & 0 & 0 & 0 & 0 & 0 & 0 & 0 & 0 & 0 & 0 & 0 & 0 & 0 & 0 & 0 & 0 & 0 & 0 & 0 & 0 & \tiny\textcircled{\emph{c}} & 0 & 0 & 0 & 0 & 0 & 0 & 0 & 0 \\
0 & 0 & 0 & 0 & 0 & 0 & 0 & b & 0 & 0 & 0 & 0 & 0 & 0 & 0 & 0 & 0 & 0 & 0 & 0 & 0 & 0 & 0 & 0 & 0 & 0 & 0 & 0 & 0 & 0 & 0 & 0 & 0 & 0 & 0 & 0 & 0 & 0 & 0 & 0 & b & 0 & 0 & 0 & 0 & 0 & 0 & 0 \\
0 & 0 & 0 & 0 & 0 & 0 & d & 0 & 0 & 0 & 0 & 0 & 0 & 0 & 0 & 0 & 0 & 0 & 0 & 0 & 0 & 0 & 0 & 0 & 0 & 0 & 0 & 0 & 0 & 0 & 0 & 0 & 0 & 0 & 0 & 0 & 0 & 0 & 0 & 0 & 0 & d & 0 & 0 & 0 & 0 & 0 & 0 \\
0 & 0 & 0 & 0 & 0 & \tiny\textcircled{\emph{c}} & 0 & 0 & 0 & 0 & 0 & 0 & 0 & 0 & 0 & 0 & 0 & 0 & 0 & 0 & 0 & 0 & 0 & 0 & 0 & 0 & 0 & 0 & 0 & 0 & 0 & 0 & 0 & 0 & 0 & 0 & 0 & 0 & 0 & 0 & 0 & 0 & \tiny\textcircled{\emph{c}} & 0 & 0 & 0 & 0 & 0 \\
0 & 0 & 0 & 0 & b & 0 & 0 & 0 & 0 & 0 & 0 & 0 & 0 & 0 & 0 & 0 & 0 & 0 & 0 & 0 & 0 & 0 & 0 & 0 & 0 & 0 & 0 & 0 & 0 & 0 & 0 & 0 & 0 & 0 & 0 & 0 & 0 & 0 & 0 & 0 & 0 & 0 & 0 & b & 0 & 0 & 0 & 0 \\
0 & 0 & 0 & c & 0 & 0 & 0 & 0 & 0 & 0 & 0 & 0 & 0 & 0 & 0 & 0 & 0 & 0 & 0 & 0 & 0 & 0 & 0 & 0 & 0 & 0 & 0 & 0 & 0 & 0 & 0 & 0 & 0 & 0 & 0 & 0 & 0 & 0 & 0 & 0 & 0 & 0 & 0 & 0 & c & 0 & 0 & 0 \\
0 & 0 & b & 0 & 0 & 0 & 0 & 0 & 0 & 0 & 0 & 0 & 0 & 0 & 0 & 0 & 0 & 0 & 0 & 0 & 0 & 0 & 0 & 0 & 0 & 0 & 0 & 0 & 0 & 0 & 0 & 0 & 0 & 0 & 0 & 0 & 0 & 0 & 0 & 0 & 0 & 0 & 0 & 0 & 0 & b & 0 & 0 \\
0 & b & 0 & 0 & 0 & 0 & 0 & 0 & 0 & 0 & 0 & 0 & 0 & 0 & 0 & 0 & 0 & 0 & 0 & 0 & 0 & 0 & 0 & 0 & 0 & 0 & 0 & 0 & 0 & 0 & 0 & 0 & 0 & 0 & 0 & 0 & 0 & 0 & 0 & 0 & 0 & 0 & 0 & 0 & 0 & 0 & b & 0 \\
a & 0 & 0 & 0 & 0 & 0 & 0 & 0 & 0 & 0 & 0 & 0 & 0 & 0 & 0 & 0 & 0 & 0 & 0 & 0 & 0 & 0 & 0 & 0 & 0 & 0 & 0 & 0 & 0 & 0 & 0 & 0 & 0 & 0 & 0 & 0 & 0 & 0 & 0 & 0 & 0 & 0 & 0 & 0 & 0 & 0 & 0 & a \\
\end{smallmatrix}$}\right),
\end{eqnarray}
where $\small\textcircled{\emph{c}}$ and $\small\textcircled{\emph{d}}$ denote two diagonal $c$ and $d$, respectively. Substituting this matrix
into the Lindblad equation leads to four coupled equations
\begin{eqnarray}
\left\{
\begin{array}{l}
\dot{a}(t) = 6k\Big(b(t)-a(t)\Big),\\
\dot{b}(t) = k\Big(a(t)-6b(t)+5c(t)\Big),\\
\dot{c}(t)=2k\Big(b(t)-3c(t)+2d(t)\Big),\\
\dot{d}(t)=-6k\Big(c(t)-d(t)\Big),\\
\end{array}\right.
\end{eqnarray}
subject to $a(0)=1/2$ and $b(0)=c(0)=d(0)=0$. Thus, the solutions
read
\begin{eqnarray}
\left\{
\begin{array}{l}
a(t) =\frac{1}{64}\left( 1 + 15 e^{-4 \kappa t} + 15 e^{-8 \kappa t} + e^{-12 \kappa t}\right),\\
b(t) =\frac{1}{64}\left( 1 +5 e^{-4 \kappa t} - 5 e^{-8 \kappa t} - e^{-12 \kappa t}\right),\\
c(t) =\frac{1}{64}\left( 1 -  e^{-4 \kappa t} -  e^{-8 \kappa t} +
e^{-12 \kappa
t}\right),\\
d(t) =\frac{1}{64}\left( 1 - 3 e^{-4 \kappa t} + 3 e^{-8 \kappa t}-
e^{-12 \kappa t}\right),
\end{array}\right.
\end{eqnarray}
and finally
\begin{eqnarray}
F(\theta, \phi) &=&  \frac{1}{2} \left[ 1 + \sin^2 \theta \cos^2
\phi+ e^{-4 \kappa t} (\cos^2 \theta +
     \sin^2 \theta \sin^2 \phi )\right],
   \\ \overline{F} &=& \frac{1}{3} \left( 2 + e^ {-4 \kappa t}
\right).
\end{eqnarray}

For the last case, we study
($L_{2,z}$,$L_{3,z}$,$L_{4,z}$,$L_{5,z}$,$L_{6,z},L_{7,z}$) noise
with
$\kappa_{2,z}=\kappa_{3,z}=\kappa_{4,z}=\kappa_{5,z}=\kappa_{6,z}=\kappa_{7,z}=\kappa$.
For this case, the temporal evolution matrix is
\begin{eqnarray}
\varepsilon(\rho_{6\mbox{\tiny GHZ}})\Big|_{t=\delta t} =\frac{1}{2}
\left( |0\rangle^{\otimes 6} \langle 0 |^{\otimes 6} + |1\rangle
^{\otimes 6}\langle 1| ^{\otimes 6}\right) + \frac{1-12\kappa\delta
t}{2} \left(|0\rangle ^{\otimes 6} \langle 1| ^{\otimes 6}+
|1\rangle ^{\otimes 6}\langle 0 |^{\otimes 6} \right),
\end{eqnarray}
Therefore, using the ansatz
\begin{eqnarray}
\varepsilon(\rho_{6\mbox{\tiny GHZ}}) = a \left( |0\rangle^{\otimes
6} \langle 0 |^{\otimes 6} + |1\rangle ^{\otimes 6}\langle 1|
^{\otimes 6}\right) + b\left(|0\rangle ^{\otimes 6} \langle 1|
^{\otimes 6}+ |1\rangle ^{\otimes 6}\langle 0 |^{\otimes 6} \right),
\end{eqnarray}
we obtain two simple differential equations
\begin{eqnarray}
\left\{
\begin{array}{l}
\dot{a}(t) =0,\\
\dot{b}(t) = -12kb(t),\\
\end{array}\right.
\end{eqnarray}
subject to $a(0)=b(0)=1/2$. So the solution is given by
\begin{equation}
\label{zmatrix6} \varepsilon(\rho_{6\mbox{\tiny GHZ}}) = \frac{1}{2}
\left( |0\rangle^{\otimes 6} \langle 0 |^{\otimes 6} + |1\rangle
^{\otimes 6}\langle 1| ^{\otimes 6}\right) + \frac{1}{2} e^{-12
\kappa t} \left(|0\rangle ^{\otimes 6} \langle 1| ^{\otimes 6}+
|1\rangle ^{\otimes 6}\langle 0 |^{\otimes 6} \right),
\end{equation}
and the fidelity and its average read
\begin{eqnarray}
F(\theta, \phi) &=&  1 - \frac{1}{2} \left( 1 - e^{-12 \kappa t}\right) \sin^2 \theta ,\\
\overline{F} &=& \frac{1}{3} \left( 2 + e^ {-12 \kappa t} \right).
\end{eqnarray}

\section{Conclusions}\label{sec6}
In this paper, we studied quantum teleportation through noisy
channels for $n$GHZ states, $n\in\{4,5,6\}$), so that the noisy
channels lead to the quantum channels to be mixed states. We
exactly solved the Lindblad equation and obtained corresponding
density matrices after the transmission process. The Lindblad
operators are responsible for the decoherence of quantum states and
are defined to be proportional to the Pauli matrices. Solving the
Lindblad equation for $n>2$ is not a trivial task in general. For
instance, we need to solve 2080 coupled differential equations to
find the density matrix for 6GHZ state. We overcame this problem by
studying the temporal evolution of the input state and using a
proper ansatz for the density matrix. Therefore, we reduced 2080
coupled equations to at most four coupled equations which are
readily solved. We found the fidelity and the average fidelity for
various cases and showed that for the Lindblad operators
corresponding to $x$ direction the fidelity is the same for EPR and
$n$GHZ states where $n\in\{3,4,5,6\}$. However, 3GHZ state does lose
less quantum information for other types of noisy channel. Note
that, In Ref.~\cite{jung08-2} the authors only studied the same-axis
noisy channels and conjectured that ``average fidelity with
same-axis noisy channels are in general larger than average fidelity
with different-axis noisy channels''. However, we showed the failure
of this conjecture for 4GHZ state which is apparent in
Fig.~\ref{fig7}. In the appendix we showed this conjecture also
fails for 3GHZ state (see Fig.~\ref{fig6}). In fact, for
different-axes noises, the analytical solutions can be obtained in
the same way, but the number of coupled differential equations
usually increases with respect to the same-axes noises.

\acknowledgments
We would like to thank Robabeh Rahimi for fruitful discussions and
suggestions and for a critical reading of the paper.

\newpage
\appendix

\section{}
\begin{figure}
\centering
\includegraphics[width=8cm]{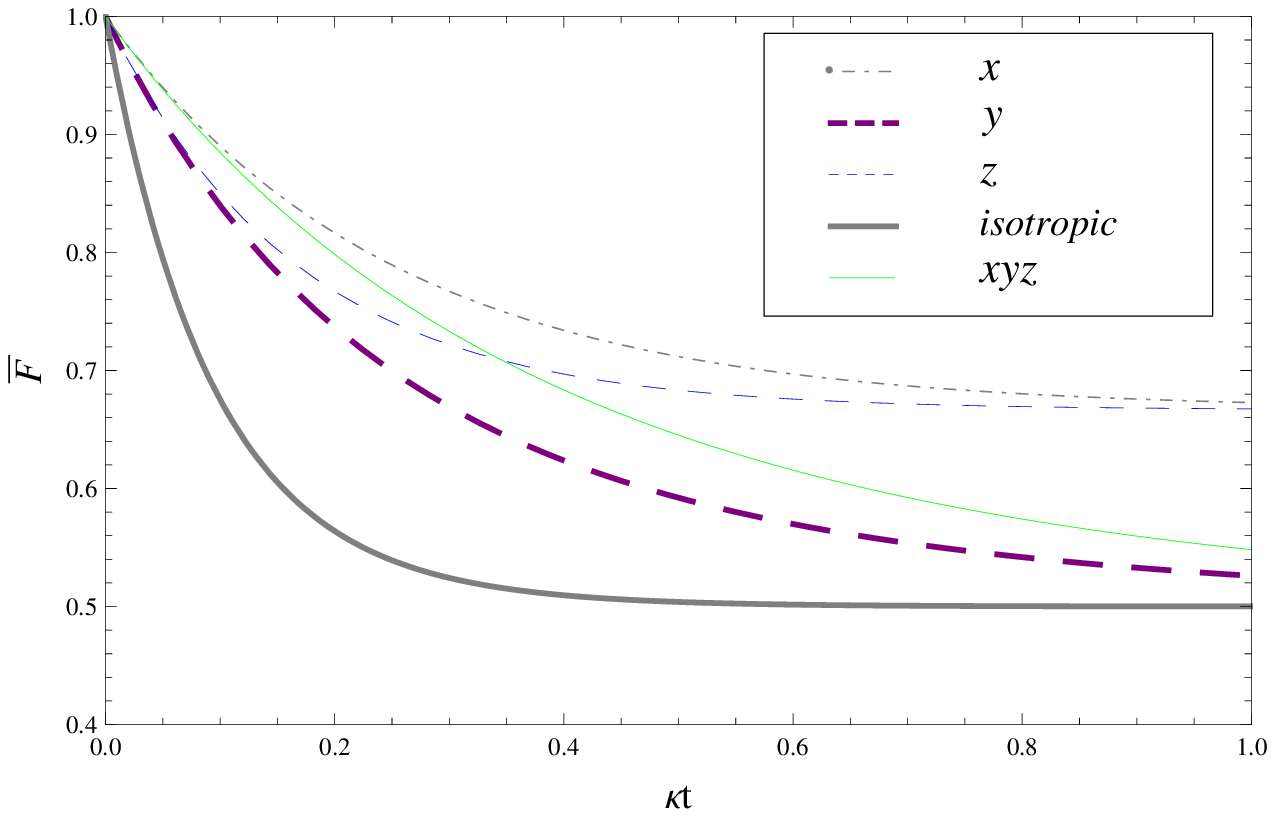}
\caption{\label{fig6} The plot of time dependence of average
fidelity for $(L_{2,x},L_{3,y},L_{4,z}$) noisy channels for 3GHZ
state.}
\end{figure}

Here, we present quantum teleportation process through
($L_{2,x},L_{3,y},L_{4,z}$) noisy channel for 3GHZ state which is
not studied in Ref.~\cite{jung08-2}. For this case, the density
matrix after $\delta t$ reads
\begin{eqnarray}
\varepsilon(\rho_{3\mbox{\tiny GHZ}})\Big|_{t=\delta t} =
\frac{1}{2}{\left(
\begin{smallmatrix}
1 -2 \kappa \delta t & 0 & 0 & 0 & 0 & 0 & 0 & 1-4 \kappa \delta t \\
0 & 0 & 0 & 0& 0 & 0 & 0 & 0 \\
0 & 0 & \kappa \delta t & 0 & 0 & -\kappa \delta t & 0 & 0 \\
0 & 0 & 0 & \kappa \delta t & \kappa \delta t & 0& 0 & 0 \\
0 & 0 & 0 & \kappa \delta t & \kappa \delta t & 0& 0 & 0 \\
0 & 0 & -\kappa \delta t & 0 & 0 & \kappa \delta t & 0 & 0 \\
0 & 0 & 0 & 0 & 0 & 0 & 0& 0 \\
1 -4 \kappa \delta t & 0 & 0 & 0 & 0
&0 & 0 & 1 -2 \kappa \delta t
\end{smallmatrix} \right).}
\end{eqnarray}
So, we examine the following ansatz
\begin{eqnarray}
\varepsilon(\rho_{3\mbox{\tiny GHZ}}) =
 \left(
\begin{smallmatrix}
a & 0 & 0 & 0 & 0 & 0 & 0 & d   \\
0 & b& 0 & 0 & 0 & 0 & e & 0   \\
0 & 0 & c & 0 & 0 & f & 0 & 0   \\
0 & 0 & 0 & c & g & 0 & 0 & 0   \\
0 & 0 & 0 & g & c & 0 & 0 & 0   \\
0 & 0 & f & 0& 0 & c & 0 & 0   \\
0 & e & 0 & 0& 0 & 0 & b & 0   \\
d & 0 & 0 & 0 & 0 & 0 & 0& a
\end{smallmatrix}
\right),
\end{eqnarray}
which results in two sets of coupled equations
\begin{eqnarray}
\left\{
\begin{array}{l}
\dot{a}(t) = 2k\Big(c(t)-a(t)\Big),\\
\dot{b}(t) = 2k\Big(c(t)-b(t)\Big),\\
\dot{c}(t) = k\Big(a(t)+b(t)-2c(t)\Big),
\end{array}\right.
\end{eqnarray}
and
\begin{eqnarray}
\left\{
\begin{array}{l}
\dot{d}(t) = k\Big(g(t)-4d(t)-f(t)\Big),\\
\dot{e}(t) = k\Big(f(t)-4e(t)-g(t)\Big),\\
\dot{f}(t) = k\Big(e(t)-d(t)-4f(t)\Big),\\
\dot{g}(t) = k\Big(d(t)-e(t)-4g(t)\Big),
\end{array}\right.
\end{eqnarray}
subject to $a(0)=d(0)=1/2$ and $b(0)=c(0)=e(0)=f(0)=g(0)=0$. The
solutions are
\begin{eqnarray}\left\{
\begin{array}{l}
a(t) =e^{2 \kappa t}d(t)=\frac{1}{8}\Big( 1 + 2 e^{-2 \kappa t} + e^{-4 \kappa t}\Big),\\
b(t) =-e^{2 \kappa t}e(t)=\frac{1}{8}\Big( 1 -2 e^{-2 \kappa t} + e^{-4 \kappa t} \Big),\\
c(t) =e^{2 \kappa t}g(t)=-e^{2 \kappa t}f(t)=\frac{1}{8}\Big( 1 -
e^{-4 \kappa t}\Big).
\end{array}\right.
\end{eqnarray}
By using the unitary gate matrix which can be read off from Fig.~2 of Ref.~\cite{jung08-2}, the fidelity, $F(\theta, \phi)$, and the average fidelity, $\overline{F}$, are given by
\begin{eqnarray}
F(\theta, \phi) = \frac{1}{2} \left[ 1 + e^{-2 \kappa
t}\left(\cos^2 \theta + \sin^2 \theta \sin^2 \phi
\right)+ e^{-4 \kappa t} \sin^2 \theta \cos^2 \phi
\right],
\end{eqnarray}
and
\begin{eqnarray}
 \overline{F} = \frac{1}{6} \left( 3 + 2e^{-2 \kappa t}+ e^{-4 \kappa t} \right).
\end{eqnarray}
In Fig.~\ref{fig6}, we depicted the average fidelity for 3GHZ state
through various noises where the results for the same-axes and
isotropic noises are given in Ref.~\cite{jung08-2}. Therefore, the
average fidelity for ($L_{2,x},L_{3,y},L_{4,z}$) noise explicitly
contradicts the conjecture proposed by Jung \emph{et al.}
\cite{jung08-2}.

\end{document}